\documentclass[fleqn,usenatbib]{mnras}

\usepackage{newtxtext,newtxmath}

\usepackage[T1]{fontenc}
\usepackage{ae,aecompl}

\usepackage{graphicx}	
\usepackage{amsmath}	
\usepackage{amssymb}	
\usepackage{pifont}
\usepackage{times}
\usepackage{ulem}
\usepackage{xcolor}

\newcommand{\cmark}{\ding{51}}
\newcommand{\xmark}{\ding{55}}
\newcommand{\vect}[1]{\boldsymbol{#1}}

\usepackage{etoolbox}

\defcitealias{PaperII}{Paper~II}

\title[HD~100453 I. The orbit of the binary]{Spirals, shadows \& precession in HD~100453 -- I. The orbit of the binary}

\author[J.-F. Gonzalez et al.]{Jean-Fran\c{c}ois Gonzalez,$^{1}$\thanks{E-mail: jean-francois.gonzalez@ens-lyon.fr}
Gerrit van der Plas,$^{2}$
Christophe Pinte,$^{2,3}$
Nicol\'as Cuello,$^{2}$
\newauthor
Rebecca Nealon,$^{4}$
Fran\c{c}ois M\'enard,$^{2}$
Alexandre Revol,$^{1}$
Laetitia Rodet,$^{5}$
\newauthor
Maud Langlois$^{1}$
and Anne-Lise Maire$^{6}$
\\
$^{1}$Univ Lyon, Univ Claude Bernard Lyon 1, ENS de Lyon, CNRS, Centre de Recherche Astrophysique de Lyon UMR5574, F-69230, Saint-Genis-Laval, France\\
$^{2}$Univ. Grenoble Alpes, CNRS, IPAG (UMR 5274), F-38000 Grenoble, France\\
$^{3}$Monash Centre for Astrophysics (MoCA) and School of Physics and Astronomy, Monash University, Clayton Vic 3800, Australia\\
$^{4}$School of Physics and Astronomy, University of Leicester, University Road, Leicester LE1 7RH, UK\\
$^{5}$Cornell Center for Astrophysics and Planetary Science, Department of Astronomy, Cornell University, Ithaca, NY 14853, USA\\
$^{6}$STAR Institute, Universit\'e de Li\`ege, All\'ee du Six Ao\^ut 19c, 4000 Li\`ege, Belgium
}

\date{Accepted 2020 September 22. Received 2020 September 14; in original form 2020 June 5}

\pubyear{2020}

\begin{document} 
\label{firstpage}
\pagerange{\pageref{firstpage}--\pageref{lastpage}}
\maketitle

\begin{abstract}
In recent years, several protoplanetary discs have been observed to exhibit spirals, both in scattered light and (sub)millimetre continuum data. The HD~100453 binary star system hosts such a disc around its primary. Previous work has argued that the spirals were caused by the gravitational interaction of the secondary, which was assumed to be on a circular orbit, coplanar with the disc (meaning here the large outer disc, as opposed to the very small inner disc). However, recent observations of the CO gas emission were found incompatible with this assumption. In this paper, we run SPH simulations of the gas and dust disc for seven orbital configurations taken from astrometric fits and compute synthetic observations from their results. Comparing to high-resolution ALMA $^{12}$CO data, we find that the best agreement is obtained for an orbit with eccentricity $e=0.32$ and semi-major axis $a=207$~au, inclined by $61\degr$ relative to the disc plane. The large misalignment between the disc and orbit planes is compatible with the tidal evolution of a circumprimary disc in an eccentric, unequal-mass binary star.
\end{abstract}




\section{Introduction}
\label{Sec:Intro}

HD~100453 is a binary star system composed of a Herbig A9Ve primary (A) of mass $M_\mathrm{A}=1.7$~M$_\odot$ \citep{Dominik2003} and a secondary (B) discovered by \citet{Chen2006}. Its common proper motion was established by \citet{Collins2009}, who derived its spectral type of M4V to M4.5V and mass $M_\mathrm{B}=0.20\pm0.04$~M$_\odot$. The \citet{GaiaDR2} measured the distance to HD~100453~A to be $104.2\pm0.4$~pc and \citet{Vioque2018} derived an age of $6.5\pm0.5$~Myr from that same data. \citet{Wagner2018} presented a collection of astrometric measurements putting HD~100453~B at a relatively stable separation over the 2003--2017 period of $\simeq1\farcs05$ and a position angle varying from $\simeq127\degr$ in 2003 to $\simeq132\degr$ in 2015--2017, yielding a projected separation of $\simeq109$~au.

The presence of a disc around HD~100453~A was first inferred from its infrared spectral energy distribution (SED) by \citet{Meeus2001}. Observing with VLT/SPHERE in scattered light, \citet{Wagner2015} detected a ring between $0\farcs18$ and $0\farcs25$ in radius, which we will refer to as the outer disc, inclined $\sim34\degr$ from face-on, with asymmetric features and surrounding a cavity. Their images also show two highly symmetric spiral arms extending out to $0\farcs37$. Near infrared interferometric observations with VLTI/MIDI \citep{Menu2015} and VLTI/PIONIER \citep{Lazareff2017,Kluska2020} revealed the presence of an inner disk, with a half light radius of 2.6~mas. In polarized scattered light at optical and near infrared wavelengths, \citet{Benisty2017} additionally detected two symmetric shadows in the outer disc, as well as a fainter spiral-like feature interpreted as emission from the surface of the bottom side of the disc. They proposed that the shadows are cast by the unresolved inner disc with inclination $i=-48\degr$ and position angle PA~=~$80\degr$, inclined by $72\degr$ with respect to the outer disc, with $i=38\degr$ and PA~=~$142\degr$ (the inner disc orientation was later measured by the interferometric observations of \citealt{Kluska2020} to $i=-44\pm5\degr$ and PA~=~$92\pm8\degr$). The same features were also seen in polarimetric imaging with the Gemini Planet Imager by \citet{Long2017}, who fitted both the SED and images to derive an outer disc inclination of $25\pm10\degr$ and position angle of $140\pm10\degr$.

\citet{vanderPlas2019} observed the HD~100453 system with ALMA in band 6. They detected the outer disc in the 1.4~mm dust continuum, extending from $0\farcs22$ to $0\farcs40$, with $i=29\fdg5\pm0\fdg5$ and PA = $151\fdg0\pm0\fdg5$. The data constrained the dust disc mass to 0.07~M$_\mathrm{J}$ and provided an upper limit on the dust mass around HD~100453~B of 0.03~M$_\oplus=9.5\times10^{-5}$~M$_\mathrm{J}$. The gas disc was also detected in the $^{12}$CO, $^{13}$CO and C$^{18}$O $J=2$--1 emission lines,  extending out to $1\farcs10$, $0\farcs70$ and $0\farcs50$, respectively, with no apparent cavity. The total gas mass estimated from CO is 0.001--0.003~M$_\odot$. Both gas and dust masses agree with upper limits from \citet{Kama2020}. Recently, \citet{Rosotti2020} presented higher-resolution ALMA band 7 data revealing spiral arms counterparts to the scattered-light ones in both the sub-mm continuum and the $^{12}$CO $J=3$--2 emission line. In the latter, the disc extends out to $0\farcs33$ and the spirals reach out to larger radii, $\sim0\farcs6$, with the southern arm connecting to the projected position of the secondary star. They measured the pitch angle of the spirals to be $\sim6\degr$ in the sub-mm continuum and $\sim16\degr$ in the R' scattered light data from \cite{Benisty2017}. They fitted their projected velocity map to obtain $i=35\degr$ and PA = $145\degr$.

Several causes have been suggested for the grand design spiral structure observed in HD~100453 A's outer disc. \citet{Benisty2017} explored the possibility that the shadows seen in polarized scattered light could trigger the spiral arms via a pressure decrease, a mechanism proposed by \citet{Montesinos2016} and \citet{Montesinos2018}. The hypothesis that the spiral arms are instead due to the tidal interaction with the secondary star has been more thoroughly investigated. \citet{Dong2016} performed hydrodynamical and radiative transfer simulations of the circumprimary disc, using $M_\mathrm{B}=0.3$~M$_\odot$, and showed that the spirals observed in scattered light can be explained by perturbations from the companion, supposed to be on a circular and coplanar orbit. They further assumed that the disc is close to face-on, which is however not supported by the observations. \citet{Wagner2018} fitted their astrometric data to determine the parameters of the companion's orbit, with $M_\mathrm{B}=0.2$~M$_\odot$. They found a semi-major axis of $a=1\farcs06\pm0\farcs09$, an eccentricity of $e=0.17\pm0.07$ and an inclination of $i=32\fdg5\pm6\fdg5$, at first sight approximately coplanar with the disc. They then used hydrodynamical and radiative transfer simulations with a coplanar, circular orbit with $a=100$~au to compute synthetic images in near infrared scattered light and again found a spiral structure qualitatively similar to the observed one.

\citet{vanderPlas2019} challenged the results of both previous works, which assumed a low-eccentricity, nearly coplanar orbit, because the gas disc extends beyond the secondary, well outside the primary's Roche lobe. They performed smoothed particle hydrodynamics (SPH) simulations of a gas-and-dust disc with the same orbital parameters as \citet{Dong2016}, which resulted in a gas disc that was too small to reproduce the spatial extent of the observed CO disc, hinting at a significant misalignment between orbit and disc. \citet{vanderPlas2019} calculated new solutions for the secondary's orbit using the same astrometric data and found that an inclined orbit is indeed favoured, with a most probable relative angle of $\sim60\degr$ (although the probability distribution is broad, see Section~\ref{Sec:MethSetup}). These are different from \cite{Wagner2018} because the latter omitted to take into account the difference in longitude of nodes when calculating the relative inclination between orbit and disc. \citet{Rosotti2020} also performed hydrodynamic simulations of the system, assuming a coplanar and circular orbit, and showed that the differing pitch angles seen in scattered light and sub-mm continuum can only be explained if there exists a temperature difference between the cold mid-plane and hot surface layers. They find that their higher resolution ALMA data do not reveal a large CO disc, hinting at a coplanar orbit in which a tidal truncation would be expected, but not favouring or excluding any specific scenario. They however note that their largest recoverable scale of 0.6~arcsec and lower sensitivity may not allow the detection of such a large disc.

Binary stars systems are natural and frequent outcomes of the star formation process. The collapse and fragmentation of molecular cloud cores \citep{Boss1979} or the fragmentation of a gravitationally unstable disc \citep{Bonnell1994} have long been the generally accepted mechanisms to account for the majority of such systems, see e.g.\ reviews by \citet{Duchene2013} or \citet{Reipurth2014}. The observed multiplicity of low-mass systems results mainly from turbulent cloud fragmentation, for which higher eccentricities and orbital misalignments with the stellar spin or discs are expected, while high-mass are better explained by disc instability \citep{Offner2010}. Simulations of star cluster formation by \citet{Bate2012} show broad distributions for semi-major axes and eccentricities and nearly flat mass ratio distributions. Binary stars can also form by capture during stellar encounters, but such events are generally considered to be very rare in the field. However, in the denser environments of star formation and early evolution, their probability can reach $\sim30\%$ \citep{Winter2018}. Due to their random nature, no strong bias in orbital elements is expected from this formation pathway. Additionally, binary systems are frequently formed in more dissipative star-disc encounters in the early stages of the fragmentation of turbulent molecular clouds \citep{Bate2018}. They naturally result in circumstellar discs that are misaligned with the binary orbit, with a preference for alignement that increases for closer binaries.

In this paper, our goal is to bring further constraints, independent of astrometric fitting, to the orbit of the secondary in the HD~100453 system using hydrodynamical simulations to reproduce the observed outer disc morphology, following the method for HD~142527 by \citet{Price2018}. In a companion paper \citep[][hereafter Paper~II]{PaperII}, we investigate the mass and location of a suspected inner companion and consider the long term evolution of the system, explaining the origin of the misaligned discs using the orbit of the outer companion from this work. Hereafter in this paper, we will not take the inner disc into consideration and refer to the outer disc simply as the disc. The paper is organised as follows: we present our methods in Section~\ref{Sec:Methods}, detail our hydrodynamical simulations and how they provide constraints on the binary orbit in Section~\ref{Sec:Constraining} and perform a more detailed comparison to observations in Section~\ref{Sec:CompObs}. We discuss our results in Section~\ref{Sec:Discussion} and conclude in Section~\ref{Sec:Concl}.

\section{Methods}
\label{Sec:Methods}

\subsection{Hydrodynamic simulations}
\label{Sec:MethHydro}

We perform global 3D simulations of the HD~100453 system with the SPH \citep[see][for reviews]{Monaghan2005,Price2012} public code \textsc{Phantom} \citep{PhantomCode,Phantom2018}. The SPH method is particularly well suited to the complex geometry of our study (see Section.~\ref{Sec:MethAngles}), with no preferred plane. We run a first set of simulations for a gas-only disc and a second set for a gas+multigrain dust disc. In the latter, the mixture of gas and 11 populations of dust grains (with sizes logarithmically spaced between 1.6~$\mu$m and 1.6~mm following a power-law size distribution of index $-3.5$) is treated with a single set of SPH particles and evolved according to the algorithms of \citet{Hutchison2018} and \citet{Ballabio2018}. Grains in each size bin experience gas drag according to their size, and their cumulative back-reaction on the gas is taken into account self-consistently. We take an intrinsic density of $\rho_\mathrm{s}=3\times10^3$~kg\,m$^{-3}$  for the solids and an initially uniform dust-to-gas mass ratio of 0.01.

In both sets of simulations we use $10^6$ SPH particles. We initially distribute them to reproduce a power-law disc around the primary star with surface density $\Sigma\propto R^{-1}$, where $R$ is the cylindrical radius, and use a locally isothermal equation of state where the sound speed is a function of the spherical radius $r$ as $c_\mathrm{s}\propto r^{-1/4}$. The disc has a mass $M_\mathrm{disc}=0.003$~M$_\odot$ and extends between $R_\mathrm{in}=12$~au and $R_\mathrm{out}=60$~au. We take a disc aspect ratio $H/R=0.05$ at $R_\mathrm{in}$, yielding $H/R\simeq0.075$ at $R_\mathrm{out}$. We choose a fixed artificial viscosity parameter $\alpha_\mathrm{AV}=0.18$ to recover an average \citet{SS1973} viscosity parameter $\alpha_\mathrm{SS}=5\times10^{-3}$, following \citet{Lodato2010}. The stars are treated as sink particles with masses $M_\mathrm{A}=1.7$~M$_\odot$ and $M_\mathrm{B}=0.2$~M$_\odot$, they interact with the disc particles via gravity and accretion \citep[see][for details]{Phantom2018}. We set their accretion radius to 12~au for the primary star and 5~au for the secondary. Since we do not focus on reproducing the inner disc in this paper, we choose a rather large value for the accretion radius of the primary, equal to $R_\mathrm{in}$, which is itself sufficiently far away from the observed inner edge of the outer disc to prevent our inner boundary conditions to affect the disc structure.
Finally, we adopt an orientation for the disc defined by its inclination $i_\mathrm{disc}=29\degr$ and position angle $\Omega_\mathrm{disc}=150\degr$.

\subsection{Problem geometry}
\label{Sec:MethAngles}

\begin{figure}
\centering
\resizebox{\hsize}{!}{
\includegraphics{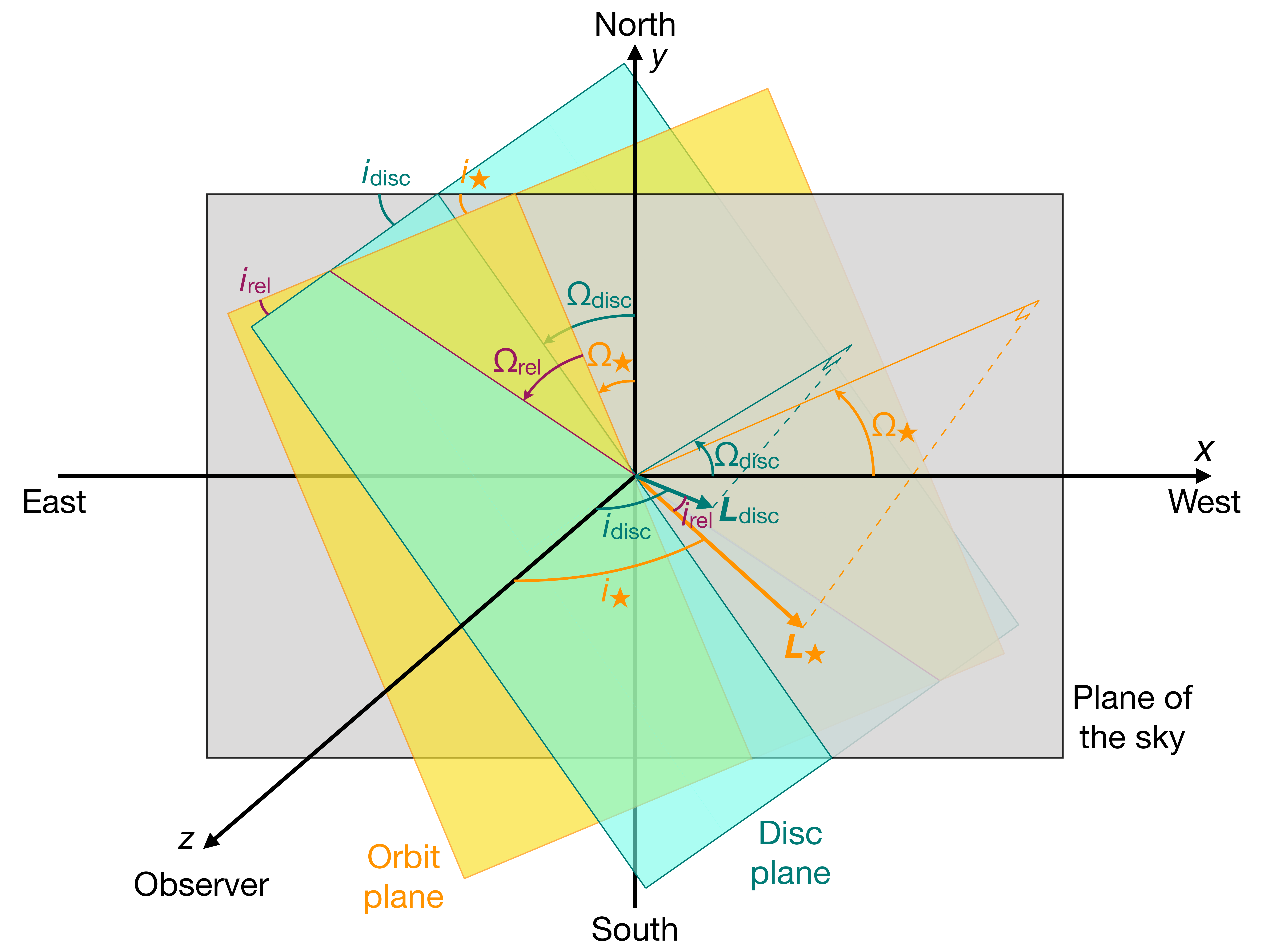}
}
\caption{Definition of the angles between the plane of the sky (grey), the orbit plane (yellow) and the disc plane (cyan). See text for details.}
\label{Fig:angles}
\end{figure}

Figure~\ref{Fig:angles} shows the geometry of the problem and defines the relevant planes and angles. Our simulations are set up so that the $xy$ plane coincides with the plane of the sky (in grey), with the $z$ axis pointing towards the observer. The plane of the binary orbit is shown in yellow and that of the disc in cyan.

Inclination (or tilt) angles of the orbit, $i_\star$, and the disc, $i_\mathrm{disc}$, are defined between their respective plane and the plane of the sky. They are equal to the angles between the perpendicular directions to each plane, i.e. between the angular momentum vector of the orbit $\vect{L}_\star$ or disc $\vect{L}_\mathrm{disc}$ and the $z$ direction. The PAs (or twist angles) of the orbit $\Omega_\star$ or disc $\Omega_\mathrm{disc}$ are the longitudes of the ascending nodes in the plane of the sky, counted counterclockwise from the North (or the $y$ direction). Similarly, they can be measured as the angles between the West (or the $x$ direction) and the projection of the angular momentum vectors on the plane of the sky, counted counterclockwise. The relative inclination between orbit and disc, $i_\mathrm{rel}$, is the angle between the orbit and disc planes, or between both angular momentum vectors $\vect{L}_\star$ and $\vect{L}_\mathrm{disc}$. The relative twist angle $\Omega_\mathrm{rel}$ is measured, after rotating the frame so that the line of nodes of the orbit plane is in the new $y$ direction (rotation of $-\Omega_\star$ about $z$) and $\vect{L}_\star$ is in the $z$ direction (rotation of $-i_\star$ about the new $y$ axis), between the new $x$ direction and the projection of $\vect{L}_\mathrm{disc}$ on the new $xy$ plane.

In our SPH simulations we use the angular momentum vectors to compute all six angles. To estimate the uncertainties on angles, we assume that the uncertainty on $\vect{L}_\star$ is negligible and that the uncertainty on $\vect{L}_\mathrm{disc}$ is the dominant contribution. $\vect{L}_\mathrm{disc}$ is perpendicular to the disc plane. The uncertainty of the orientation of the disc plane is constrained by the orientations of its upper and lower surfaces and any warp across the disc. We thus take the initial disc opening angle, defined as $\arctan(H/R)$ where $H$ is the scale-height, as a measure of the uncertainty on the orientation of $\vect{L}_\mathrm{disc}$, and therefore as the uncertainty on the disc and relative angles.

\subsection{Setup for the binary orbit}
\label{Sec:MethSetup}

\citet{vanderPlas2019} performed an astrometric fit of the binary orbit using the Markov-chain Monte Carlo (MCMC) Bayesian analysis technique \citep{Ford2005,Ford2006}. Despite the available data spanning about 1 per cent of the orbit, they obtained consistent fits with values of the reduced $\chi^2$, $\chi_\mathrm{red}^2$, between 0.5 and 2.\footnote{$\chi^2$ is computed in the usual way, i.e. the sum of the square of the distances between data and model normalised by the uncertainty, via
\begin{equation}
\chi^2 = \sum_i\frac{(x_i-x(t_i))^2}{\sigma_{x,i}^2}+\sum_i\frac{(y_i-y(t_i))^2}{\sigma_{y,i}^2},
\label{Eq:chi2}
\end{equation}
where $x_i,y_i$ are the observed coordinates on the plane of the sky, $x(t_i),y(t_i)$ the calculated coordinates for a given orbit at epoch $t_i$ and $\sigma_{x,i},\sigma_{y,i}$ the associated uncertainties. $\chi_\mathrm{red}^2$ is obtained by dividing $\chi^2$ by $N-M$, where $N=12$ is the number of data points (2 coordinates times 6 epochs)
and $M=6$ is the number of independent parameters in the fit, i.e. the 6 orbital elements.}
Their resulting probability distribution for the semi-major axis taken individually peaks around 100~au, close to the projected separation between both stars. Similarly, the likelihood for the eccentricity peaks close to zero, favouring a circular orbit. Most previous modelling work adopted values close to these. However, the probability distributions allow a wide range of values and the overall best fitting orbit (i.e. when all orbital parameters are fitted simultaneously instead of separately) has a semi-major axis $a=207$~au and an eccentricity $e=0.32$, both in the 95\% interval around the peak. Its relative inclination to the disc plane $i_\mathrm{rel}=61\degr$, computed according to
\begin{equation}
\cos i_\mathrm{rel} = \cos i_\star \cos i_\mathrm{disc} + \sin i_\star \sin i_\mathrm{disc} \cos(\Omega_\star - \Omega_\mathrm{disc}),
\label{Eq:i_rel}
\end{equation}
is in the 68\% interval. Note that its probability distribution is broad, with close to half the plausible values below $40\degr$ \citep[see Fig.~9 in][]{vanderPlas2019}, making orbits with values of $i_\mathrm{rel}\sim10$--$20\degr$ possible as well.

\begin{figure}
\centering
\resizebox{\hsize}{!}{
\includegraphics{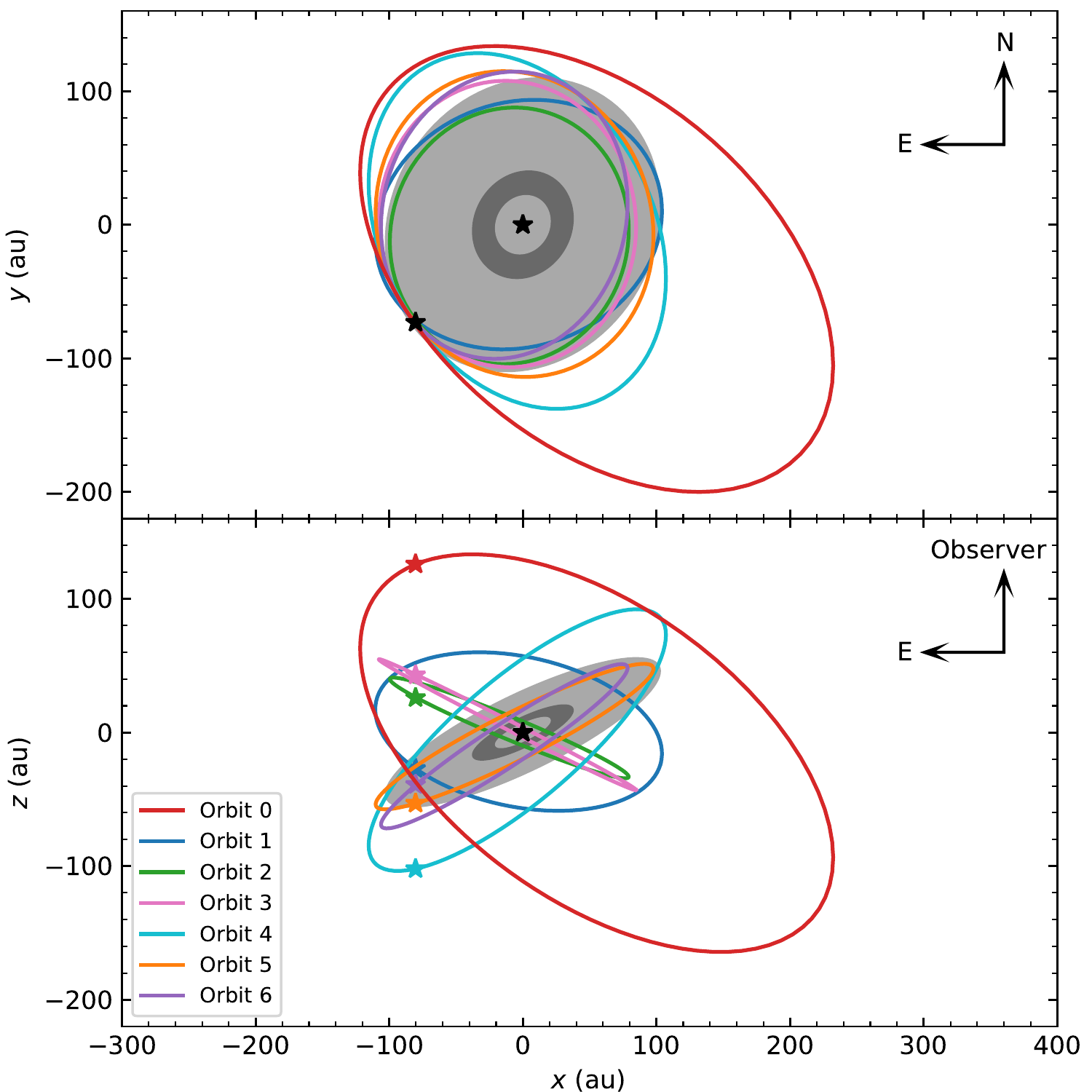}
}
\caption{Projection of the 7 orbits considered in this paper on the plane of the sky $xy$ (top panel) with North on top and East to the left, and on the perpendicular plane $xz$. The coordinate system is centered on the primary. Both stellar components are represented by a star symbol. The $^{12}$CO gas disc is shown in light grey and the dust disc in dark grey, with sizes taken from \citet{vanderPlas2019}.}
\label{Fig:orbits}
\end{figure}

Seeking to narrow down the parameter space in an independent manner, using the observed disc morphology, we selected 7 orbital solutions for our SPH simulations. We first include the overall best fitting orbit, with $\chi_\mathrm{red}^2=0.33$, which we label orbit~0. We then pick 3 orbits with semi-major axes close to the probability peak at $\sim100$~au with values of $\chi_\mathrm{red}^2 < 1$ (labelled 1, 2 and 3). These first 4 orbits all have values of $i_\mathrm{rel}$ in the $\sim50-60\degr$ interval. We continue with 2 orbits for which $i_\mathrm{rel}$ is constrained to $<40\degr$, with semi-major axes and eccentricities within the 68\% confidence interval, i.e.\ $90<a<150$~au and $e<0.2$, with $\chi_\mathrm{red}^2<1$ (labelled 4 and 5). Finally, we take an orbit as close as possible to the best fit of \citet{Wagner2018}. Since these authors did not cite values for the longitude of ascending node and argument of periastron, we select an orbit with the following constraints: $105<a<115$~au, $0.15<e<0.2$, $30<i<40\degr$ and $i_\mathrm{rel}<10\degr$ (labelled 6). The orbital parameters of our 7 orbits are listed in Table~\ref{Table:orbits}. Figure~\ref{Fig:orbits} represents the 7 orbits in the plane of the sky $xy$ and in the perpendicular plane $xz$.

\begin{table}
\caption{Orbital parameters of the 7 considered binary orbits: semi-major axis $a$, eccentricity $e$, inclination $i_\star$, longitude of ascending node $\Omega_\star$, argument of periastron $\omega$, epoch of periastron $T_\mathrm{P}$, followed by the reduced $\chi^2$ and the relative inclination $i_\mathrm{rel}$ between orbit and disc planes.}
\label{Table:orbits}
\centering
\begin{tabular}{lrrrrrrrr}
\hline
Orbit & \multicolumn{1}{c}{$a$}  & \multicolumn{1}{c}{$e$} & \multicolumn{1}{c}{$i_\star$} &  \multicolumn{1}{c}{$\Omega_\star$} &\multicolumn{1}{c}{$\omega$} & \multicolumn{1}{c}{$T_\mathrm{P}$} & \multicolumn{1}{c}{$\chi^2_\mathrm{red}$} & \multicolumn{1}{c@{}}{$i_\mathrm{rel}$}\\
& \multicolumn{1}{c}{(au)} && \multicolumn{1}{c}{($\degr$)} & \multicolumn{1}{c}{($\degr$)} & \multicolumn{1}{c}{($\degr$)} & \multicolumn{1}{c}{(yr)} && \multicolumn{1}{c@{}}{($\degr$)} \\
\hline
0 & 207 & 0.32 & 49 &   47 &  18 & 1790 & 0.33 & 61 \\
1 & 109 & 0.03 & 33 &  -71 & -24 & 2324 & 0.37 & 58 \\
2 &  97 & 0.14 & 23 &  -13 & -40 & 1725 & 0.37 & 51 \\
3 & 108 & 0.12 & 27 &    4 & -96 & 2310 & 0.56 & 53 \\
4 & 141 & 0.06 & 44 & -152 & 105 & 1425 & 0.49 & 36 \\
5 & 116 & 0.06 & 28 & -161 &  69 & 2354 & 0.86 & 23 \\
6 & 110 & 0.17 & 34 &  165 &  80 & 2248 & 0.86 &  9 \\
\hline
\end{tabular}
\end{table}

\subsection{Radiative transfer}
\label{Sec:MethRT}

We use the radiative transfer code \textsc{mcfost} \citep{Pinte2006,Pinte2009} to produce synthetic observations from the results of our gas+multigrain dust SPH simulations. The \textsc{mcfost} grid is built by performing a Voronoi tesselation with one cell per SPH particle. The dust temperature structure is computed assuming passive heating and local thermodynamic equilibrium. We set the gas temperature to be equal to the dust temperature. Both stars are assumed to be spherical and radiate isotropically as blackbodies, with $T_\mathrm{A}=7\,250$~K, $L_\mathrm{A}=6.2$~L$_\odot$ \citep{Vioque2018} and $T_\mathrm{B}=3\,250$~K, $L_\mathrm{B}\sim0.06$~L$_\odot$ \citep{Collins2009}. We compute the temperature and images using $10^8$ and $10^7$ photon packets, respectively. The dust properties are computed according to the Mie theory, assuming compact grains with an astrosilicate composition \citep{Weingartner2000}. Our grain size distribution ranges from $0.03\ \mu$m to 3~mm over 100 logarithmic bins. In each grid cell, we interpolate the density of a given grain size between the SPH grain sizes. We assume that grains smaller than $1~\mu$m, follow the gas distribution, while those larger than the largest SPH grain size, 1.6~mm, follow the distribution of the 1.6~mm grains. The overall size distribution is normalised by integrating over all grain sizes, assuming a power-law $\mathrm{d}n(a)\propto a^{-3.5}\mathrm{d}a$, and over all grid cells, such that the total dust-to-gas mass ratio is equal to 0.01. We set the $^{12}$CO abundance to a uniform value of $10^{-4}$ relative to H$_2$. We take into account CO freeze-out where $T<20$~K and photo-dissociation and photo-desorption where the ultraviolet radiation is high, following Appendix~B of \citet{Pinte2018a}. We compute scattered light images in the $I'$~band at $\lambda=0.79\ \mu$m, thermal emission at $\lambda=1.3$~mm, and $^{12}$CO $J=3$--2 molecular emission. We produce channel maps at 0.042~km\,s$^{-1}$ resolution with a turbulent velocity of 0.05~km\,s$^{-1}$, Hanning smoothed to match the observed spectral resolution. All images are convolved with a Gaussian beam matching either the angular resolution of the VLT/SPHERE observations of \citet{Benisty2017} or the ALMA CLEAN beam of the data presented in Section~\ref{Sec:ALMAdata}.

\section{Constraining the orbit}
\label{Sec:Constraining}

\subsection{``Forward'' simulations}
\label{Sec:Forward}

\begin{figure*}
\centering
\resizebox{\hsize}{!}{
\includegraphics{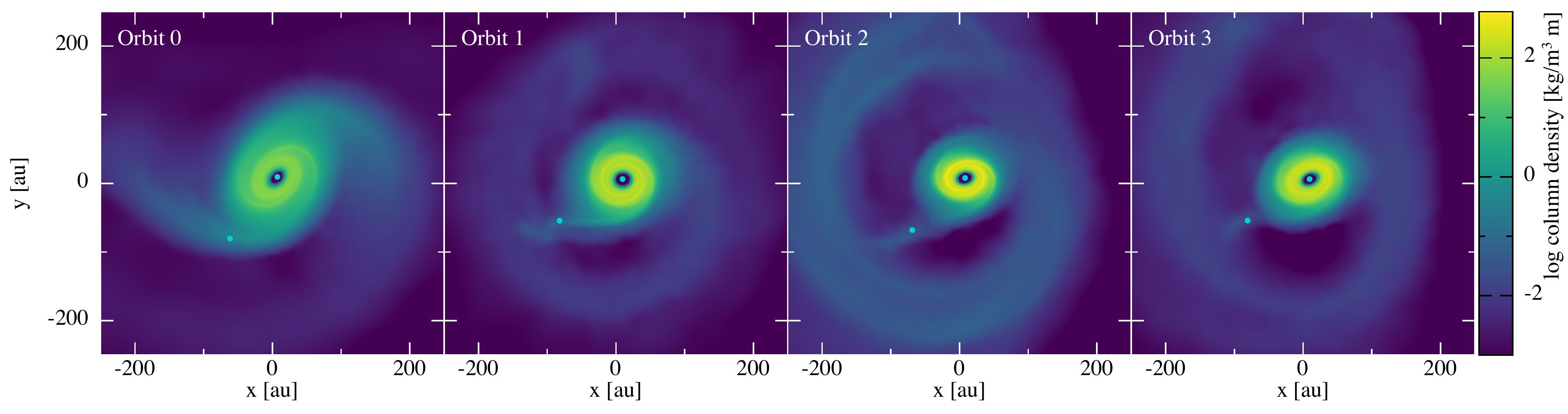}
}
\resizebox{\hsize}{!}{
\includegraphics{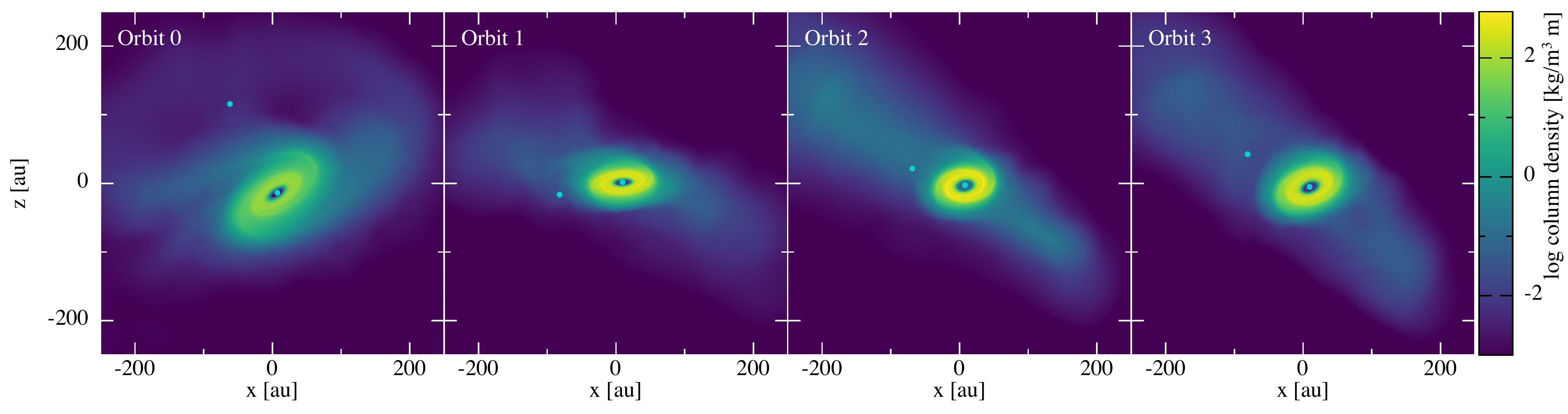}
}
\resizebox{\hsize}{!}{
\includegraphics{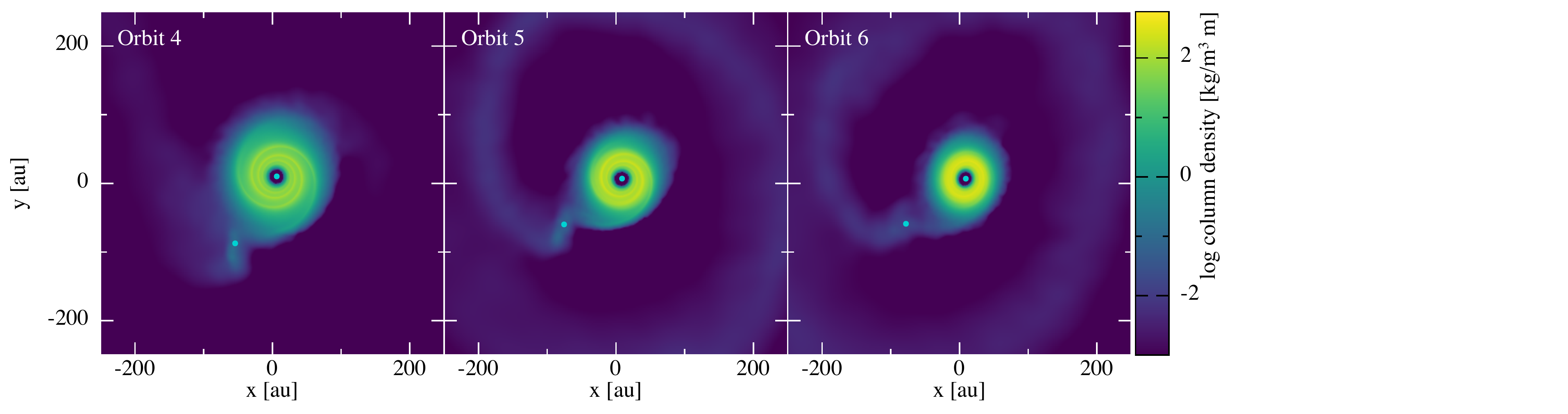}
}
\resizebox{\hsize}{!}{
\includegraphics{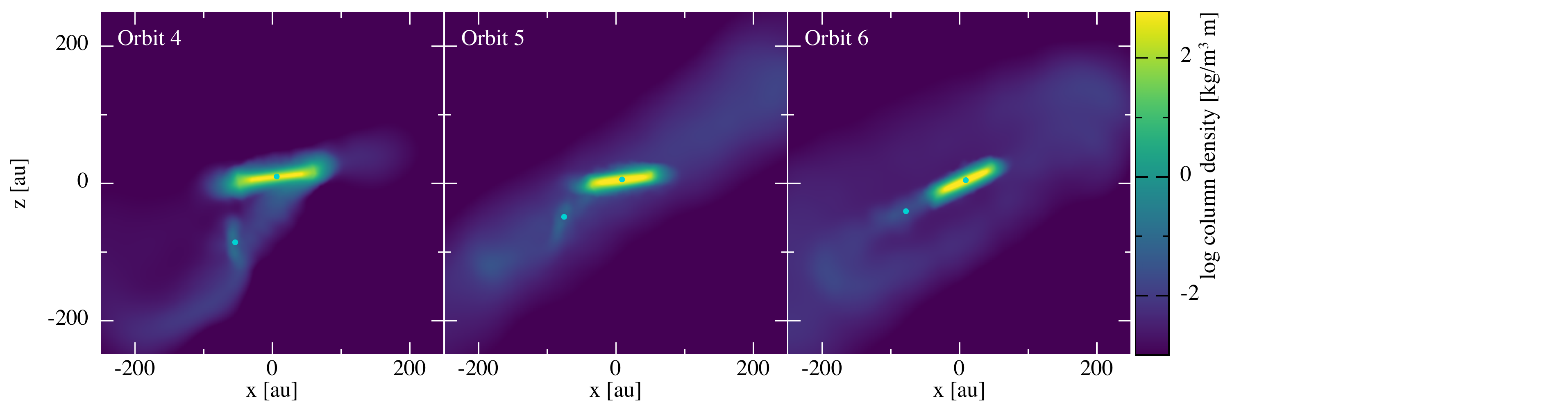}
}
\caption{Rendered column density maps of the disc in the plane of the sky $xy$ (top) and in the perpendicular plane $xz$ (bottom) after 10 orbits of the forward simulations for our 7 configurations (from left to right).}
\label{Fig:Forward10}
\end{figure*}

We start with gas-only simulations of the system for each of the four orbital configurations, with the disc initially oriented as it is observed (see Section~\ref{Sec:MethHydro}), and which we run for 100 binary orbits. We call them the ``forward'' simulations F0 to F6. Similarly to \citet{Dong2016} or \citet{Rosotti2020}, we find that HD~100453~B excites a 2-arm spiral structure that is well established within a few orbits and reaches a steady state in less than 10 orbits. Figure~\ref{Fig:Forward10} displays column density maps of the disc in the $xy$ and $xz$ planes after 10 orbits for each of our 7 configurations. Video~1, available online, shows their evolution over the first 10 orbits. The structure within the disc does not evolve significantly for longer times. In all cases, a spiral arm connects with the companion. For orbits 1, 2 and 3, which are close to circular and highly inclined, a circumbinary ring forms from the gas initially outside of the primary's Roche lobe and not captured by the secondary, while the disc remains compact. For orbits 5 and 6, having a low relative inclination, some gas is captured in a circumsecondary disc and the circumbinary ring is thinner. The disc is compact as well, with its smallest extension for orbit 6, which is close to coplanar. Orbit 4, with intermediate values of the semi-major axis and relative inclination, results in a larger disc and some circumsecondary material, but no circumbinary ring. Orbit 0 is considerably more eccentric, and highly inclined, leading to a more extended disc due to the smaller amount of time the companion spends close to the disc outer edge (see Fig.~\ref{Fig:orbits}), and to more open spiral arms. At first glance, it produces the structures most closely resembling the observations. However, the disc precesses in all configurations and, even after 10 orbits, its orientation has significantly changed from the observed one (compare Fig.~\ref{Fig:Forward10} with the sketch in Fig.~\ref{Fig:orbits} showing the initial disc orientation of the forward simulations), making the comparison more difficult.

\begin{figure}
\centering
\resizebox{\hsize}{!}{
\includegraphics{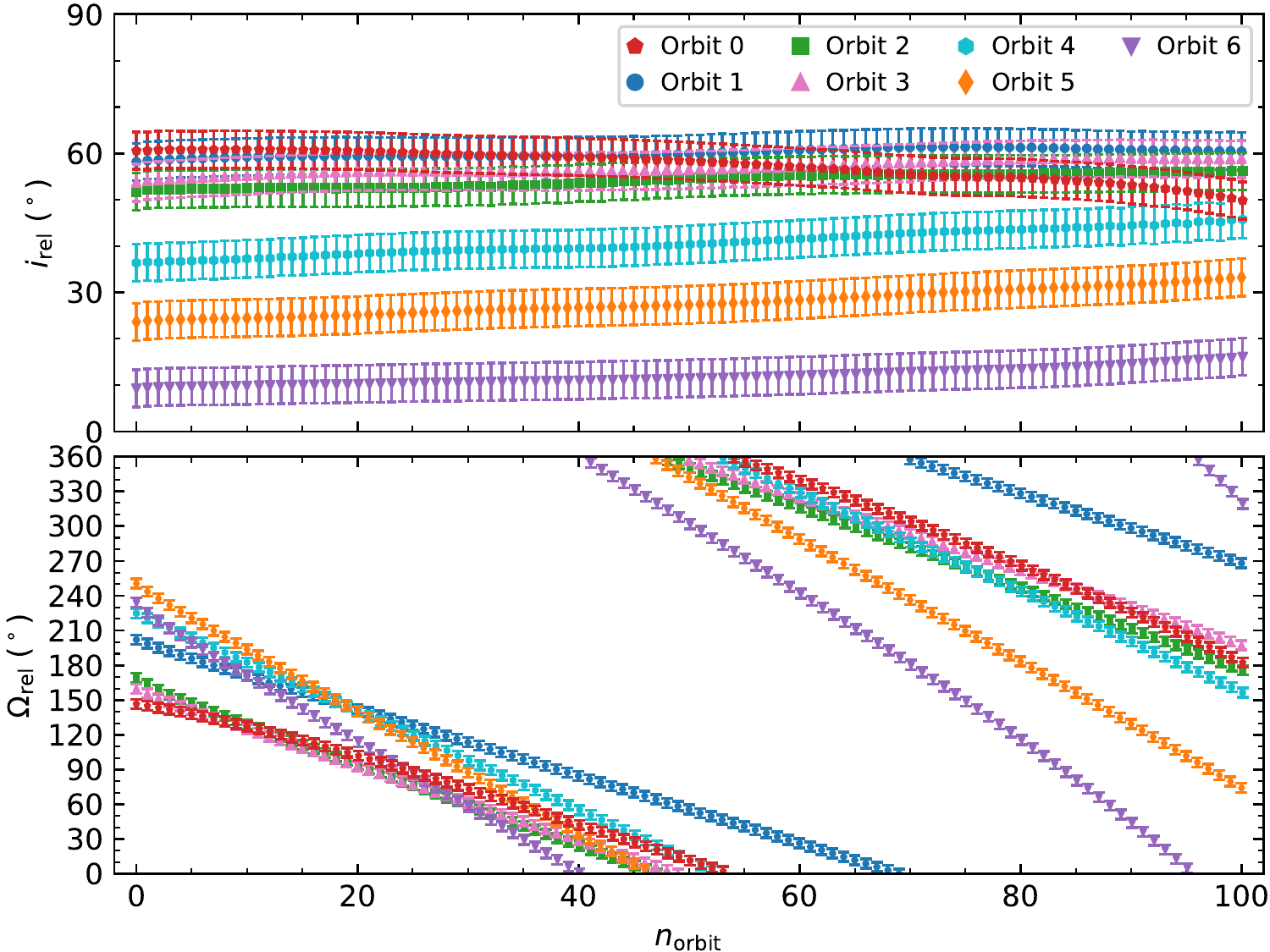}
}
\caption{Evolution of the relative tilt and twist angles over 100 orbits for our 7 orbital configurations in the forward simulations.}
\label{Fig:evol100orb}
\end{figure}

\begin{table}
\caption{Disc orientation in the gas-only simulations. F denotes the forward simulations (Section~\ref{Sec:Forward}) and R the rewind ones (Section~\ref{Sec:Rewind}) while the following digit is the orbit number.}
\label{Table:sims}
\centering
\begin{tabular}{lrrcrr}
\hline
& \multicolumn{2}{c}{Inital} && \multicolumn{2}{c@{}}{After 10 orbits} \\ \cline{2-3} \cline{5-6}
Simulation & \multicolumn{1}{c}{$i_\mathrm{disc}$} & \multicolumn{1}{c}{$\Omega_\mathrm{disc}$} && \multicolumn{1}{c}{$i_\mathrm{disc}$} & \multicolumn{1}{c@{}}{$\Omega_\mathrm{disc}$} \\
& \multicolumn{1}{c}{($\degr$)} & \multicolumn{1}{c}{($\degr$)} && \multicolumn{1}{c}{($\degr$)} & \multicolumn{1}{c@{}}{($\degr$)} \\
\hline
F0 & 29 & 150 && $44\pm5$ & $135\pm5$ \\
F1 & 29 & 150 && $27\pm4$ &  $92\pm4$ \\
F2 & 29 & 150 && $41\pm5$ & $97\pm5$ \\
F3 & 29 & 150 && $44\pm5$ & $107\pm5$ \\
F4 & 29 & 150 && $70\pm4$ & $196\pm4$ \\
F5 & 29 & 150 && $71\pm4$ & $146\pm4$ \\
F6 & 29 & 150 && $24\pm4$ & $169\pm4$ \\ \hline
R0 & 16 & 179 && $29\pm4$ & $150\pm4$ \\
R1 & 43 & 187 && $29\pm4$ & $147\pm4$ \\
R2 & 32 & 208 && $31\pm4$ & $146\pm4$ \\
R3 & 27 & 209 && $30\pm4$ & $146\pm4$ \\
R4 & 53 & 161 && $29\pm4$ & $150\pm4$ \\
R5 & 45 & 173 && $29\pm4$ & $148\pm4$ \\
R6 & 38 & 152 && $29\pm4$ & $150\pm4$ \\ \hline
\end{tabular}
\end{table}

We measure the angles $i_\star$, $\Omega_\star$, $i_\mathrm{disc}$, $\Omega_\mathrm{disc}$, $i_\mathrm{rel}$ and $\Omega_\mathrm{rel}$ as explained in Section~\ref{Sec:MethAngles}. As expected from the tidal perturbations exerted by the companion \citep[e.g.][]{Papaloizou1995}, the disc's orientation relative to the binary orbit varies notably: it tilts only slightly but it precesses with a period of the order of 100 orbits, as seen in Fig.~\ref{Fig:evol100orb} (the variation timescales of these angles are discussed in Section~\ref{Sec:DiscuMisalign}). Due to the small disc mass compared to both stellar masses, $i_\star$ and $\Omega_\star$ do not show any appreciable change in 100 orbits, but the disc orientation relative to the plane of the sky varies as well, as noted in Fig.~\ref{Fig:Forward10}. The disc inclination $i_\mathrm{disc}$ and position angle $\Omega_\mathrm{disc}$ at the beginning of the simulation and after 10 orbits are listed in Table~\ref{Table:sims} for the 7 orbital configurations.

\subsection{``Rewind'' simulations}
\label{Sec:Rewind}

In order to discriminate between the different orbital configurations, the output of our simulations should be compared to the observations when the discs have similar orientations. One possibility would be to run the simulations for one full precession period from the observed orientation. This is not practical because the slight change in inclination would modify the disc orientation relative to the plane of the sky. Instead, we choose to extrapolate the evolution of the relative angles $i_\mathrm{rel}$ and $\Omega_\mathrm{rel}$ between disc and binary orbit backwards in time to compute their values 10 orbits before the present time and run new simulations from that time onward for 10 orbits. We call these the ``rewind'' simulations R0 to R6.

\begin{table}
\caption{Linear fit parameters for the relative angles between orbit and disc in the forward simulations.}
\label{Table:fit}
\centering
\begin{tabular}{lcccc}
\hline
Simulation & \multicolumn{1}{c}{$i_\mathrm{rel}(t=0)$} & \multicolumn{1}{c}{$\displaystyle\frac{\mathrm{d}i_\mathrm{rel}}{\mathrm{d}t}$} & \multicolumn{1}{c}{$\Omega_\mathrm{rel}(t=0)$} & \multicolumn{1}{c@{}}{$\displaystyle\frac{\mathrm{d}\Omega_\mathrm{rel}}{\mathrm{d}t}$} \\
& \multicolumn{1}{c}{($\degr$)} & \multicolumn{1}{c}{($\degr/\mathrm{orbit}$)} & \multicolumn{1}{c}{($\degr$)} & \multicolumn{1}{c@{}}{($\degr/\mathrm{orbit}$)} \\
\hline
F0 & $61\pm2$ & $0.0\pm0.4$ & $147\pm2$ & $-1.9\pm0.4$ \\
F1 & $58\pm2$ & $0.1\pm0.4$ & $202\pm2$ & $-3.1\pm0.4$ \\
F2 & $52\pm2$ & $0.0\pm0.4$ & $167\pm2$ & $-3.9\pm0.4$ \\
F3 & $54\pm3$ & $0.1\pm0.4$ & $159\pm3$ & $-3.5\pm0.4$ \\
F4 & $36\pm2$ & $0.1\pm0.4$ & $225\pm2$ & $-4.2\pm0.4$ \\
F5 & $24\pm2$ & $0.1\pm0.4$ & $249\pm2$ & $-5.6\pm0.4$ \\
F6 & $10\pm2$ & $0.0\pm0.4$ & $232\pm2$ & $-6.2\pm0.4$ \\ \hline
\end{tabular}
\end{table}

\begin{figure}
\centering
\resizebox{\hsize}{!}{
\includegraphics{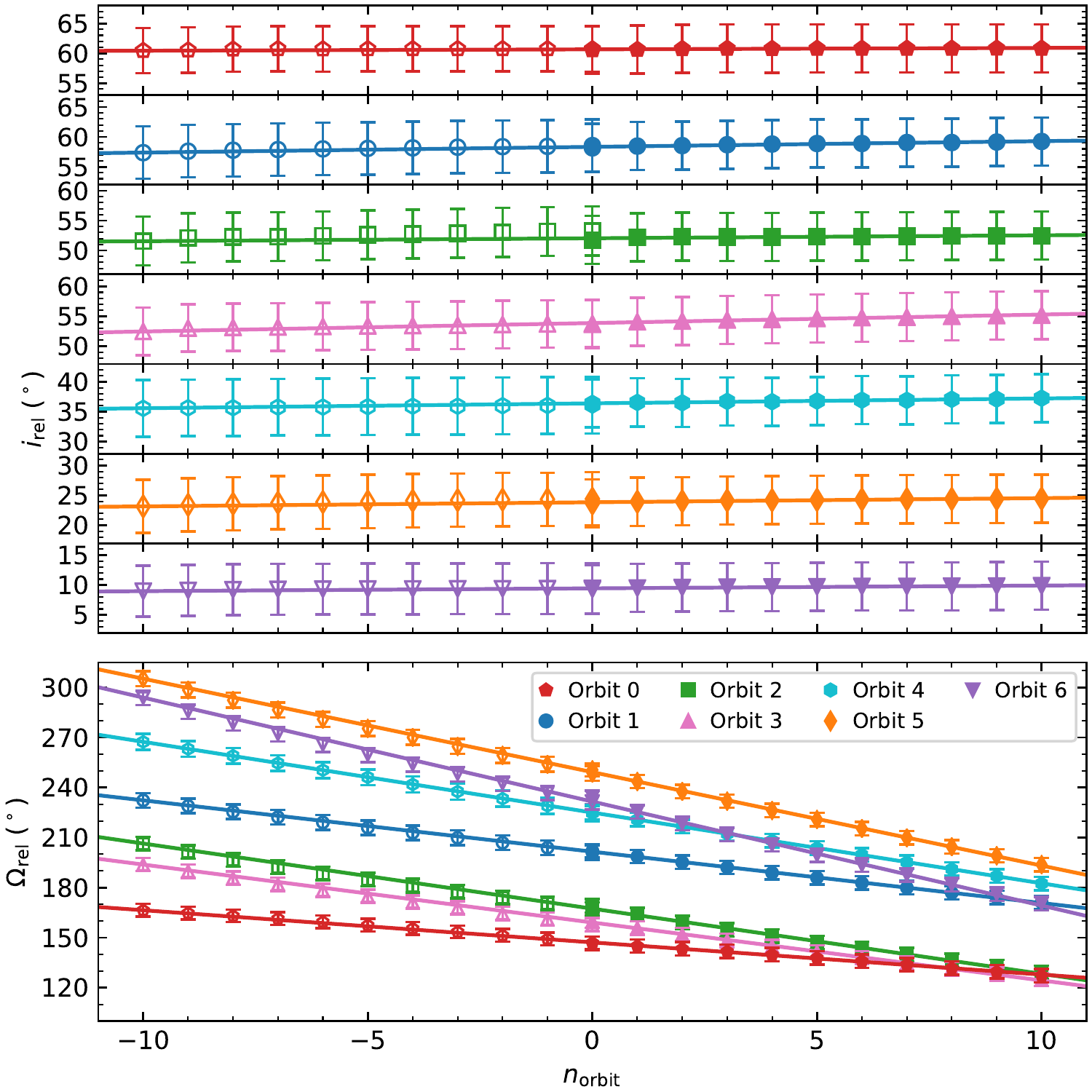}
}
\caption{Evolution of the relative tilt and twist angles for the 7 orbital configurations in the rewind simulations (open symbols) compared to the forward (filled symbols) ones. The solid lines are linear fits to the values in the forward simulations.}
\label{Fig:evol10orb_prec}
\end{figure}

To that effect, we perform both linear (of the form $a_1 + a_2t$) and non-linear (of the form $a_1 + a_2t + a_3\cos(2\pi(t-a_4)/a_5)$) fits to the values of $i_\mathrm{rel}$ and $\Omega_\mathrm{rel}$ calculated in the forward simulations, over 10 or 100 orbits. We find little difference between the various fits, with the linear fit over 10 orbits resulting in the best agreement of the disc orientation after 10 orbits of the rewind simulations with that currently observed. We thus adopt the simpler linear fit over 10 orbits, its parameters are listed in Table~\ref{Table:fit} for the seven orbital configurations. Figure~\ref{Fig:evol10orb_prec} shows the measured relative angles in both forward and rewind simulations compared to the linear fits. The evolution of both angles in the rewind simulations is in remarkable agreement with the linear fit, with the final values after 10 orbits very similar to the observed ones. The corresponding adopted initial values and final measured values of the disc inclination $i_\mathrm{disc}$ and position angle $\Omega_\mathrm{disc}$ with respect to the plane of the sky are given in Table~\ref{Table:sims}.

\begin{figure*}
\centering
\resizebox{\hsize}{!}{
\includegraphics{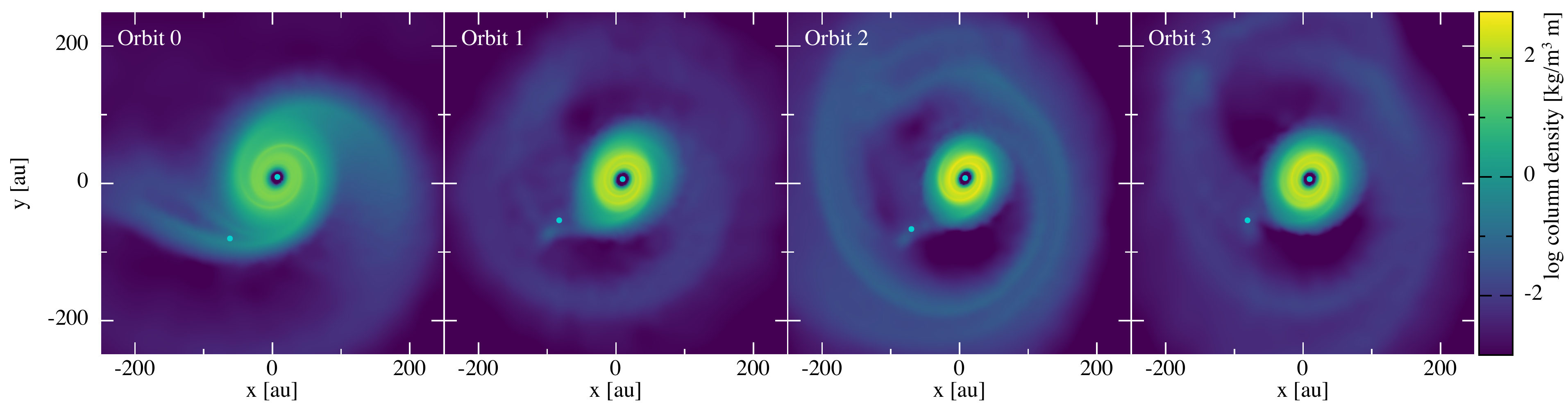}
}
\resizebox{\hsize}{!}{
\includegraphics{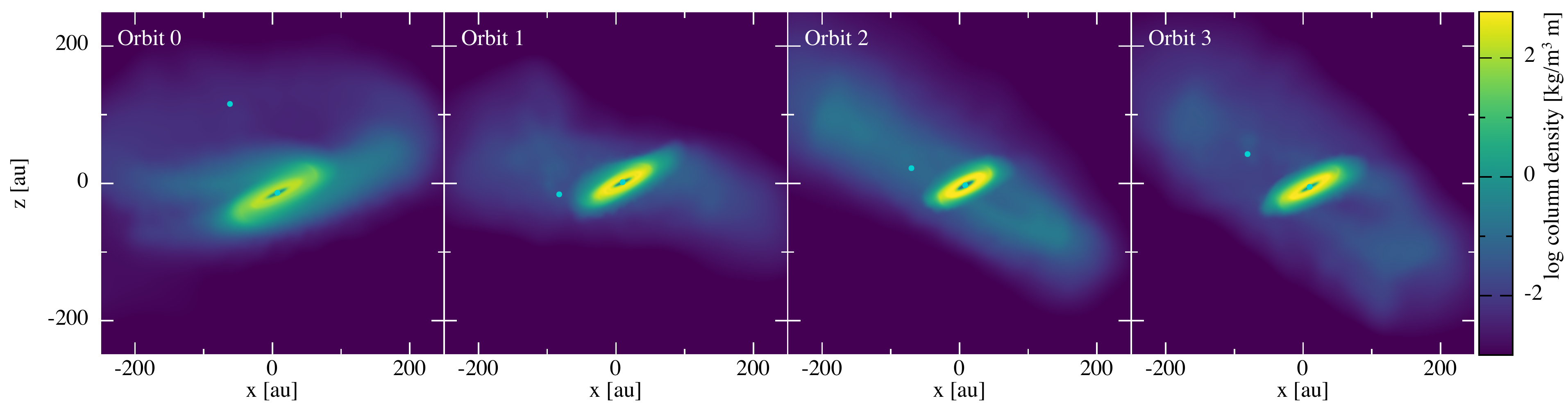}
}
\resizebox{\hsize}{!}{
\includegraphics{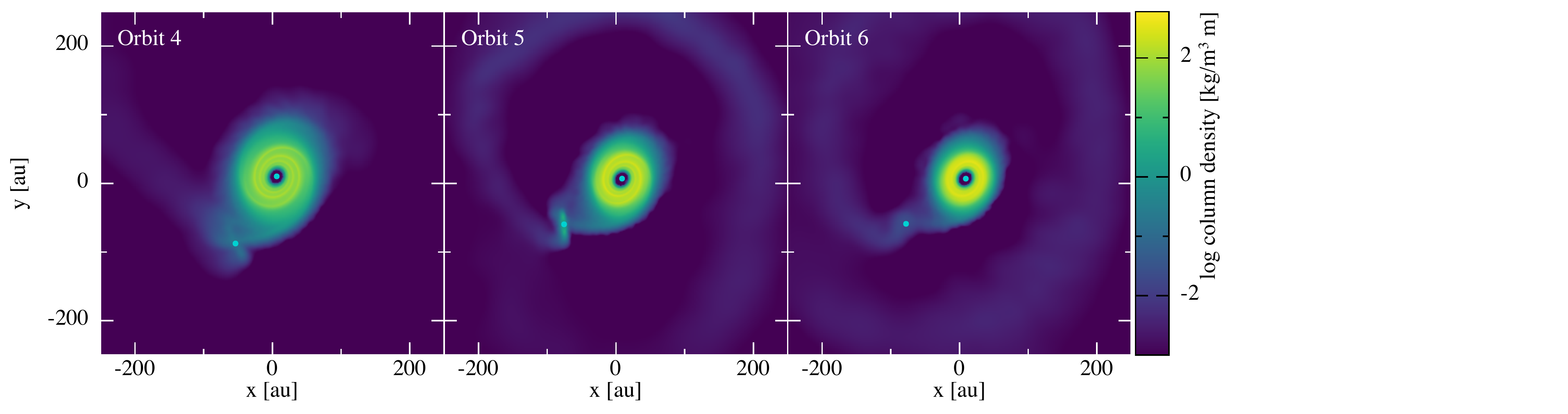}
}
\resizebox{\hsize}{!}{
\includegraphics{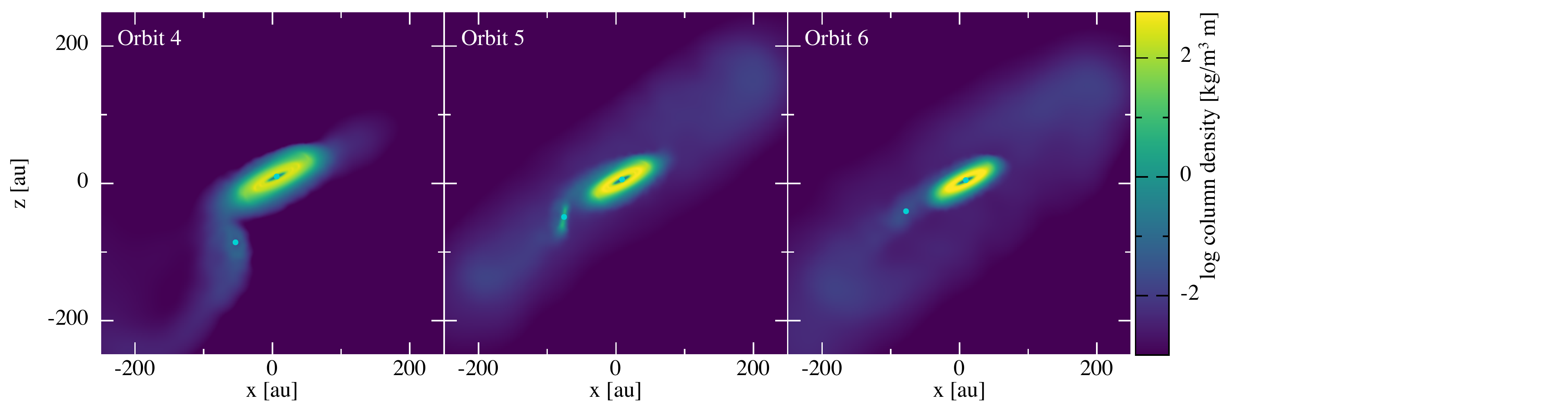}
}
\caption{Rendered column density maps of the disc in the plane of the sky $xy$ (top) and in the perpendicular plane $xz$ (bottom) after 10 orbits of the rewind simulations for our 7 configurations (from left to right).}
\label{Fig:Rewind10}
\end{figure*}

Figure~\ref{Fig:Rewind10} shows rendered column density maps of the disc in the $xy$ and $xz$ planes after 10 orbits of the rewind simulations for each of our 7 configurations, with Video~2, available online, showing their time evolution. As expected, the structures in each simulation are very similar to the ones displayed in Fig.~\ref{Fig:Forward10}, only the disc orientations are different. They are now almost identical to the observed one, as sketched in Fig.~\ref{Fig:orbits}. Here again, the structures caused by the secondary star in orbit 0 most closely resemble the observations. The spiral arms for orbit 4 seem a reasonable match as well.

\subsection{Final gas+dust simulations}
\label{Sec:Final}

\begin{figure*}
\centering
\resizebox{\hsize}{!}{
\includegraphics{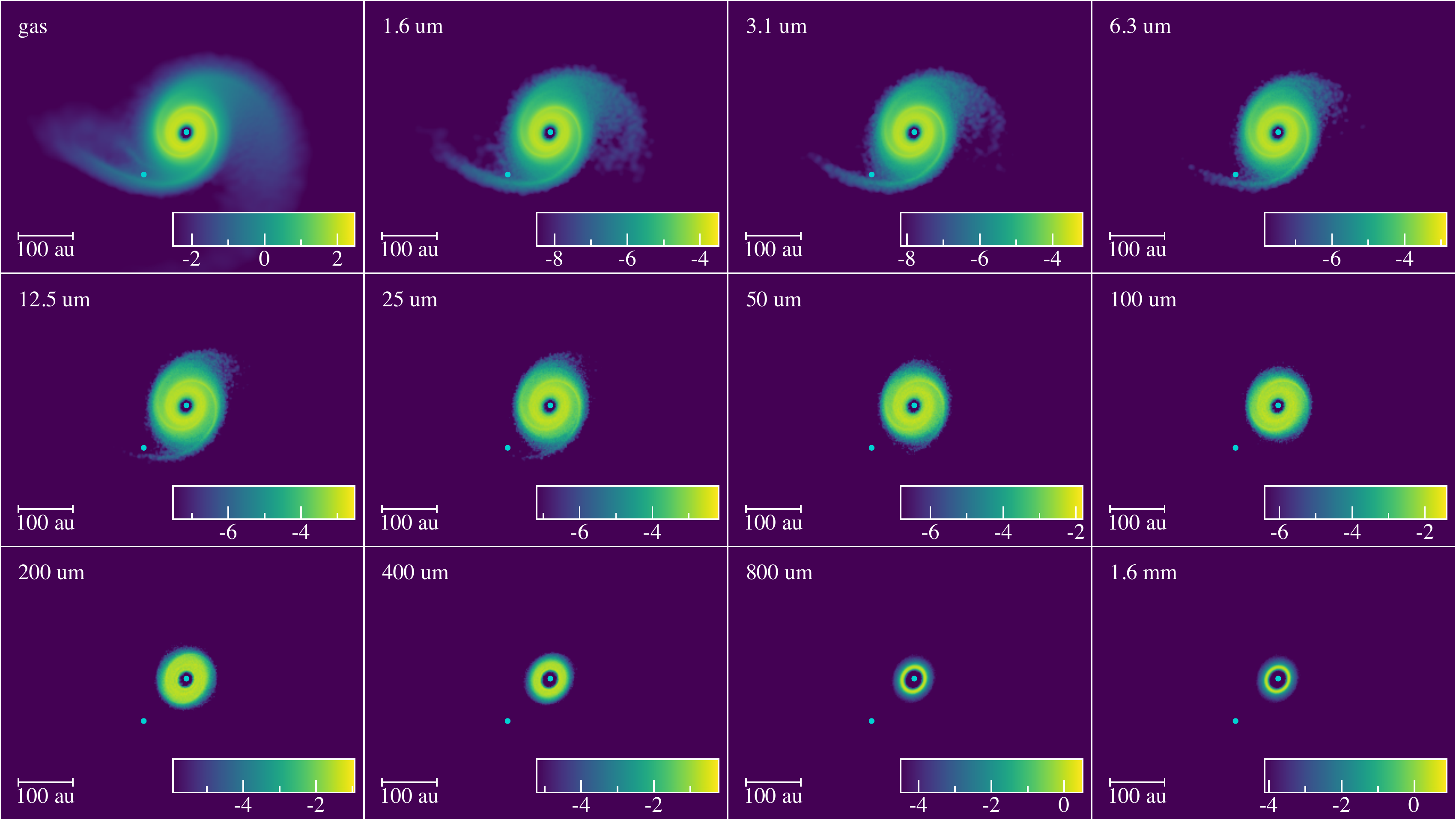}
}
\caption{Rendered column density maps of the disc in the plane of the sky after 10 orbits of the final gas+dust simulation for orbit 0, for gas and the 11 grain sizes (from left to right and top to bottom). The colourbar shows the logarithm of the column density in kg\,m$^{-2}$.}
\label{Fig:GDorb0}
\end{figure*}

A more accurate comparison with observations, done in Section~\ref{Sec:CompObs}, requires the computation of synthetic images, which themselves need the dust distribution in the system. In our final set of simulations, we now compute the coupled evolution of gas and dust (see Section~\ref{Sec:MethHydro}) from the same initial disc orientations as in the rewind simulations and for 10 orbits. The dust mass being equal to 1 per cent of that of the gas in the disc, the tidal interaction between binary orbit and disc is not affected much and the final disc orientations are almost identical to those found in the rewind simulations. Figure~\ref{Fig:GDorb0} shows rendered column density maps of the gas and each of the 11 grain sizes in the plane of the sky at the end of the simulation for orbit 0. Video~3, available online, shows their evolution. (The corresponding maps for orbits 1 to 6 are shown in Appendix~\ref{App:FigGDsims}.) The smallest grains are well coupled to the gas and are distributed similarly, while the largest ones have drifted inwards and are concentrated in a narrow, almost axisymmetric, ring. A smooth transition can be seen in between. The final state of each simulation is then fed into \textsc{mcfost} (see Section~\ref{Sec:MethRT}) to produce the synthetic observations presented in Section~\ref{Sec:CompObs}.

\section{Comparison to observations}
\label{Sec:CompObs}

\subsection{ALMA data}
\label{Sec:ALMAdata}

\begin{figure*}
\centering
\resizebox{\hsize}{!}{\includegraphics{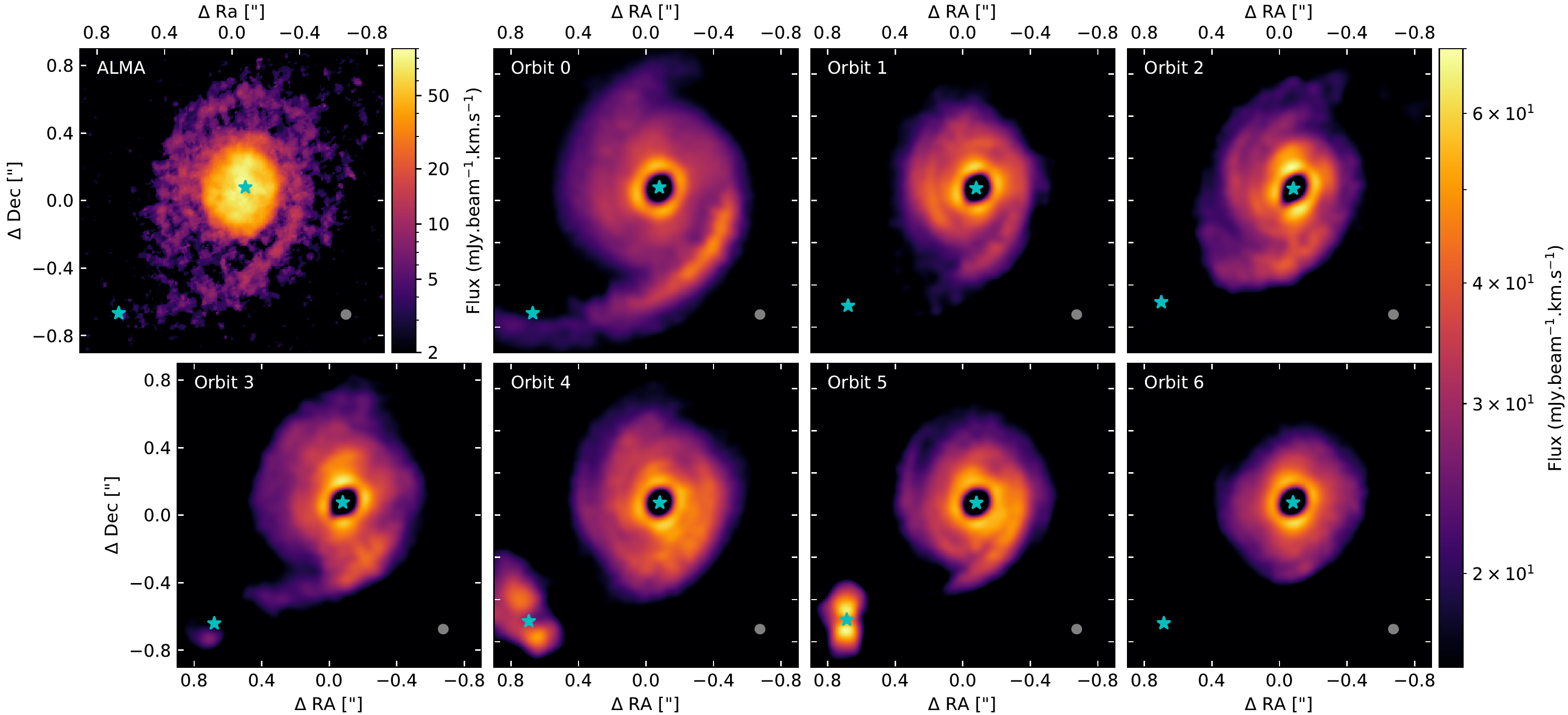}}
\caption{Moment 0 map of the $^{12}$CO $J=3$--2 line. Top left: ALMA observations, then from left to right and top to bottom: synthetic observations for orbits 0 to 6. The positions of both stars are marked in cyan and the $54\times52$~mas beam with PA~=~$83\degr$ is in the lower-right corner.}
\label{Fig:CO3-2mom0}
\end{figure*}

We use data originating from ALMA programs 2017.1.01678 (band 6, PI G. van der Plas) for the continuum image and 2017.1.01424.s (band 7, PI A. Juh\'asz) for the $^{12}$CO $J=3$--2 transition, and refer the reader to \citet{Rosotti2020} for details on the data calibration and reduction of the latter dataset. The top left panel of Fig~\ref{Fig:CO3-2mom0} shows the moment 0 map of the $^{12}$CO $J=3$--2 line.

Observations for Program 2017.1.01678 were conducted on October 21$^{\mathrm{st}}$ and on November 3$^{\mathrm{rd}}$ 2017 in the C43-10 configuration reaching a total of 92.68 minutes on source with baseline distances between 41.4 and 16196.3 meters. Note this is a different, higher spatial resolution data set than the one presented in \citet{vanderPlas2019}. During the observations the precipitable water vapor had a median value at zenith of between 0.53 and 1.04 mm.

Two of the four spectral windows of the ALMA correlator were configured in Time Division Mode (TDM) to maximise the sensitivity for continuum observations (128 channels over 1.875~GHz usable bandwidth). These two TDM spectral windows were centered at 234.18~GHz and 217.24~GHz. The data were calibrated and combined using the \textit{Common Astronomy Software Applications} pipeline \citep[CASA, ][version 5.1.1]{CASA2007}. 

We applied one round of phase-only self-calibration and imaged the continuum visibilities with the CLEAN task in CASA \citep{Hogbom1974} using Briggs weighting with a Robust parameter of 0.5, which resulted in a restored beam size of  $0\farcs025 \times 0\farcs023$ at PA~=~$2\degr$ and a RMS of 14~$\mu$Jy/beam. We show the resulting continuum map in the top left panel of Fig.~\ref{Fig:img1300}. 

\subsection{Synthetic CO gas observations}
\label{Sec:SynthGasObs}

We focus our comparison to observations of the $^{12}$CO gas emission, which can be traced to larger radii and reveals the large-scale disc structure. This is the main tracer on which we rely to draw our conclusions.

\subsubsection{Moment 0 maps}
\label{Sec:Mom0Maps}

Fig.~\ref{Fig:CO3-2mom0} shows the synthetic moment 0 maps of the $^{12}$CO $J=3$--2 line for the seven orbital configurations, to be compared to the ALMA observations shown in top left panel. Only orbit~0 produces features that are compatible with the ALMA data: the southwest spiral arm extends all the way to the secondary star with the correct orientation, and is brighter than its northeast counterpart. For orbit~1, the southwest spiral arm points towards the secondary but is too faint and short, and the northeast arm is too bright in the 4 o'clock position compared to the data. For orbits~2 and 3, the opening angle of the southwest spiral is too small and it does not point to the secondary. The spirals arms are event shorter for orbits~4 and 5, and barely discernible for orbit~6. For orbits~4 and 5, additional features not visible in the ALMA data include an eastern arm, more open than the northeast one (it may appear as a continuation of the southwest spiral for orbit~4 but this seems unlikely from Fig.~\ref{Fig:Rewind10}), and strong circumsecondary emission. The overall extent of the CO gas disc for orbit~0 is also the closest to the observed one, while it is smaller for orbits~3 and 4. It appears even more compact for the other orbits.

\begin{figure*}
\centering
\resizebox{.49\hsize}{!}{\includegraphics{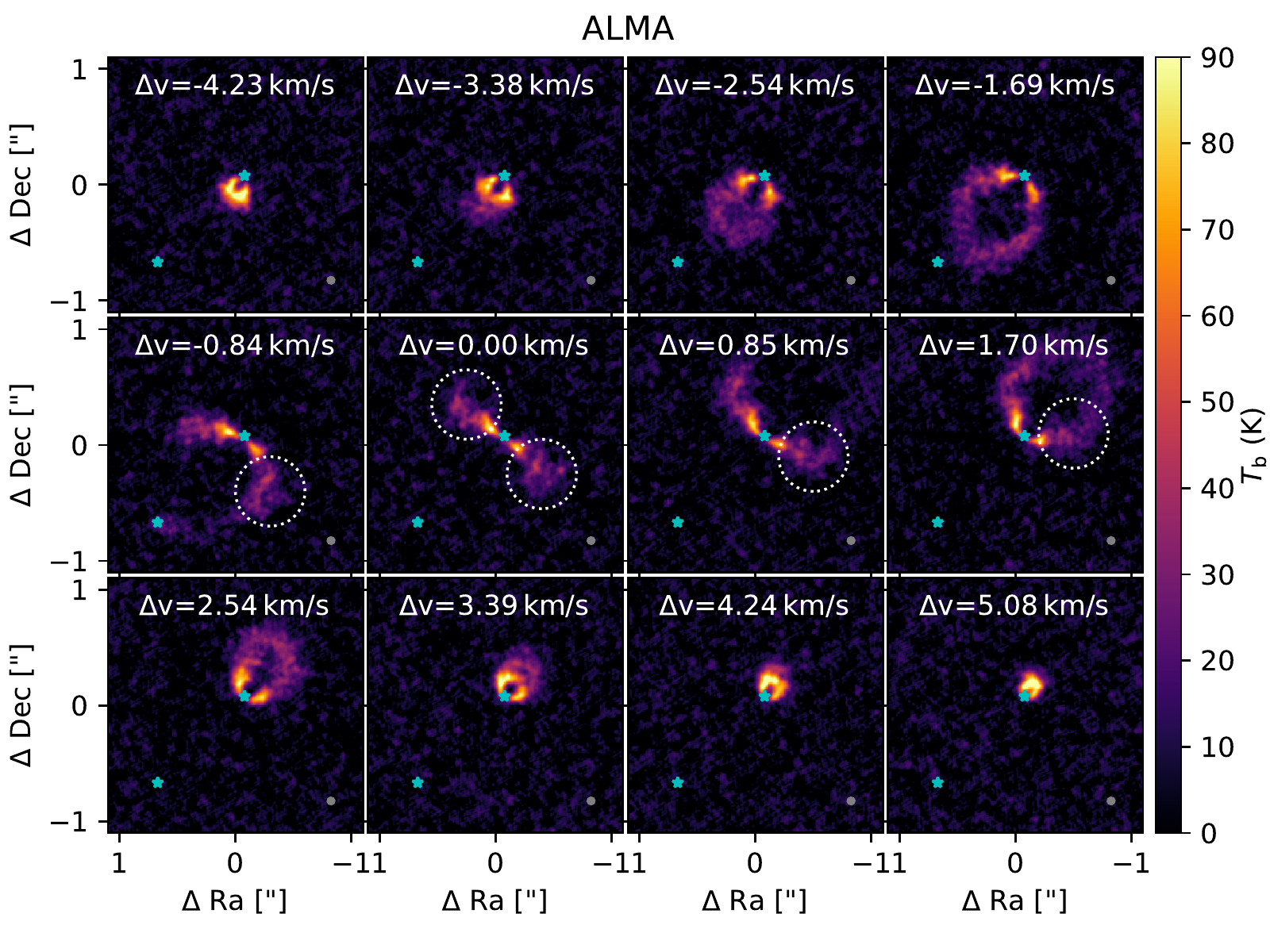}}
\resizebox{.49\hsize}{!}{\includegraphics{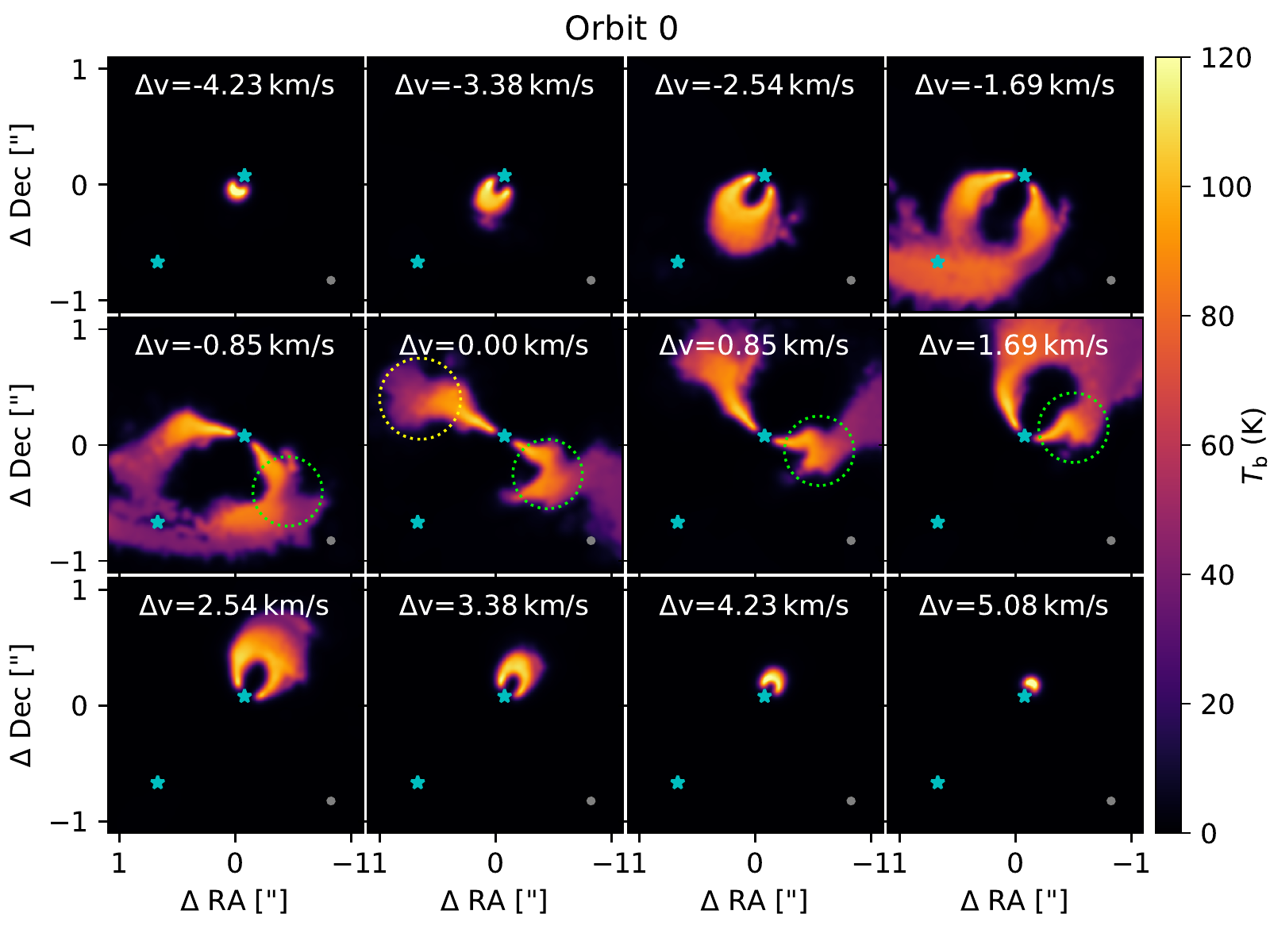}}
\caption{Channel maps of the $^{12}$CO $J =3$--2 line. Left: ALMA observations, right: synthetic observations for orbit 0. The positions of both stars are marked in cyan and the beam is in the lower-right corner. Dotted circles show the locations of the kinks in the velocity field caused by the spirals in different channels. Velocities are relative to the systemic velocity of 5.12~km\,s$^{-1}$.}
\label{Fig:CO3-2cm_obs_orb0}
\end{figure*}

\subsubsection{Channel maps}
\label{Sec:ChannelMaps}
Although so far orbit 0 is providing a closer match to the data, line emission channel maps can bring additional kinematic information to help discriminate further between the seven orbital configurations. Indeed, the spiral arms launched by HD~100453~B in the disc cause deviations in its Keplerian velocity pattern. Such `velocity kinks' are expected to be detectable even for weaker planet-induced spirals in high spectral resolution ALMA data \citep{Perez2015,Veronesi2020} and were indeed observed in several discs \citep{Pinte2018b,Pinte2019,Pinte2020}.

\begin{figure*}
\centering
\resizebox{\hsize}{!}{\includegraphics{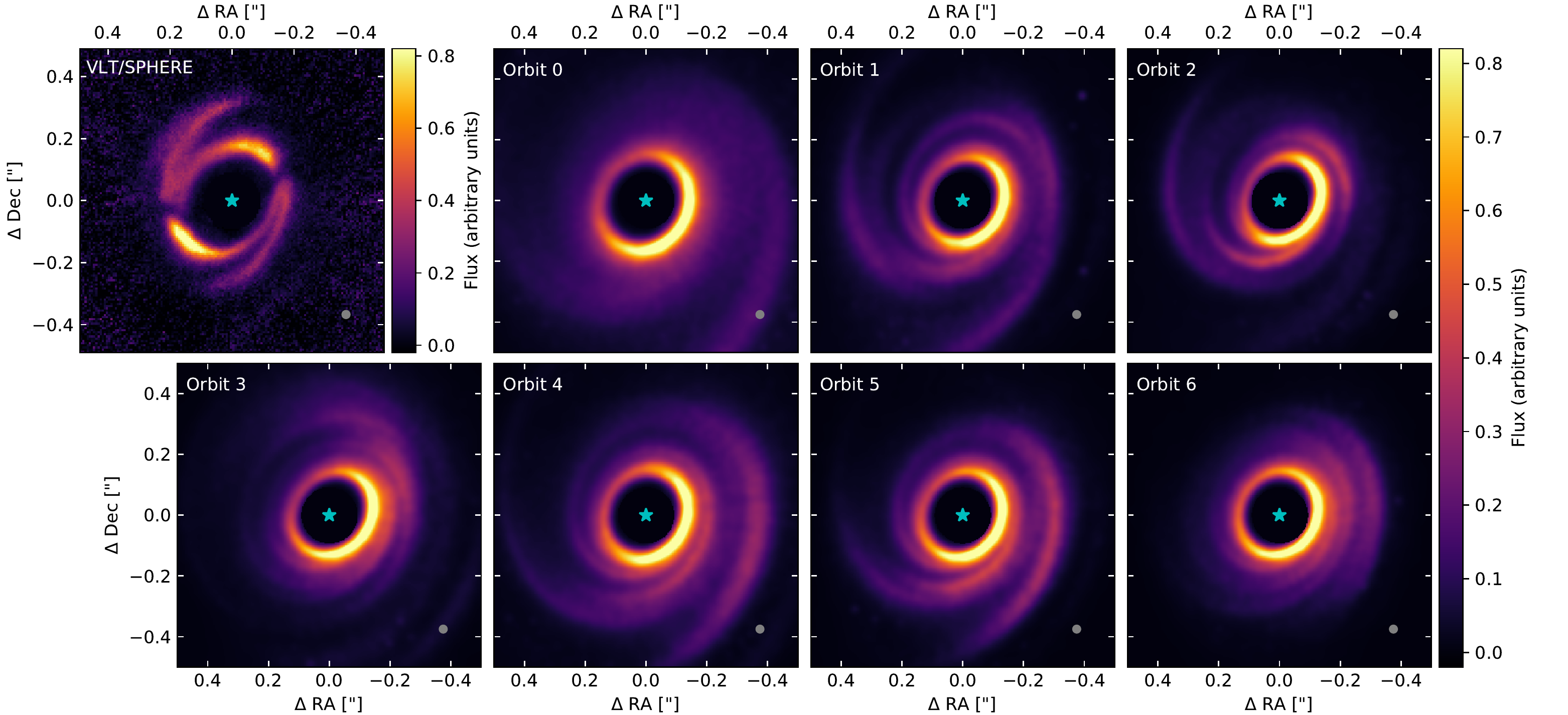}}
\caption{$I'$ band scattered light images. Top left: VLT/SPHERE observation from \citet{Benisty2017}, then from left to right and top to bottom: synthetic observations for orbits~0 to 6. The synthetic images are convolved with a Gaussian beam of $\theta=24$~mas (shown in the lower-right corner), and scaled by $r^2$, the square of the distance to the central star. We added a 185~mas-diameter coronagraph mask to hide the emission from the central star (whose position is marked in cyan), as in the VLT/SPHERE observations. Note the difference in scale with Fig.~\ref{Fig:CO3-2mom0}.}
\label{Fig:img0.79}
\end{figure*}

Although they do not reproduce every detail present in the observed channel maps of the $^{12}$CO $J=3$--2 transition (Figure~\ref{Fig:CO3-2cm_obs_orb0}, left panel), the synthetic maps computed for orbit 0 (right panel) are in broad agreement, showing the butterfly pattern typical of discs in Keplerian rotation (note that the synthetic maps do not contain any noise). In addition, the southwest spiral causes a strong kink seen across several velocity channels, which is qualitatively similar in the ALMA data and in the synthetic observations, both in location and shape. This is particularly visible in channels from $\Delta v=-0.84$ to $+1.70$~km\,s$^{-1}$ (see the dotted circles in Fig.~\ref{Fig:CO3-2cm_obs_orb0}). The northeast spiral arm causes a weaker kink more easily seen in the $\Delta v=0$ channel, seemingly farther from the central star in the synthetic map than in the observed one.

The synthetic channel maps for orbit~1 (Fig.~\ref{Fig:CO3-2cm_orb1}) are also in qualitative agreement with the ALMA data. The southwest kink has a smaller amplitude and is closer to the star than for orbit~0, while the northeast kink is stronger and visible across more channels. In the $\Delta v=0$ channel, it seems to create a disturbance towards the north while the ALMA data shows a deviation towards the south. The southwest kink for orbits~2 to 5 (Figs.~\ref{Fig:CO3-2cm_orb2} to Fig.~\ref{Fig:CO3-2cm_orb5}) has the wrong shape and for orbits 2 and 3, it even distorts the isovelocity curve in the $\Delta v=-1.69$~km\,s$^{-1}$ channel, in a different direction for each orbit. The northeast kink is stronger and points to the wrong direction for orbits~2, 4 and 5, while it is hardly detectable for orbit~3. Finally, orbit~6 does not show any appreciable kink in the southwest and only a weak one in the northeast, again pointing in the wrong direction.

The channel maps comparison again points to orbit~0 being the best candidate, although orbit~1 is harder to rule out from this criterion alone.

\subsection{Synthetic dust observations}
\label{Sec:SynthDustObs}

As explained in Section~\ref{Sec:MethHydro}, the inner radius we adopt for our SPH simulations is smaller than the observed inner edge of the disc. As such, we cannot reproduce the features of this region, which is best seen in observations of the dust phase. Indeed, scattered light observations trace the small grain population, well-coupled to the gas, at the disc surface and mostly reveal the brightest inner parts \citep[see e.g.][]{Benisty2017}. The dust continuum in the (sub)millimeter is emitted by large grains close to the mid-plane, subject to efficient radial drift and concentrated close the inner edge (see Fig.~\ref{Fig:GDorb0}). Since they are closer to the star in our simulations, the innermost grains are hotter and brighter than in the observations, and intercept photons that can no longer reach the disc immediately behind. In addition, the inclined orbits cause a warp in the disc (see \citetalias{PaperII} for details), whose illumination pattern can be affected by the radius of the inner rim. Our synthetic dust observations are thus unable to reproduce accurately the absolute or relative brightness profiles in this region and cannot be used as primary criteria to discriminate between our seven orbital configurations. Nevertheless, they still correctly trace the morphology of disc features, in particular spirals, and they exhibit differences that we use to rule out less likely candidates rather than confirm more likely ones. We show and discuss them in the following subsections.

\begin{figure*}
\centering
\resizebox{\hsize}{!}{\includegraphics{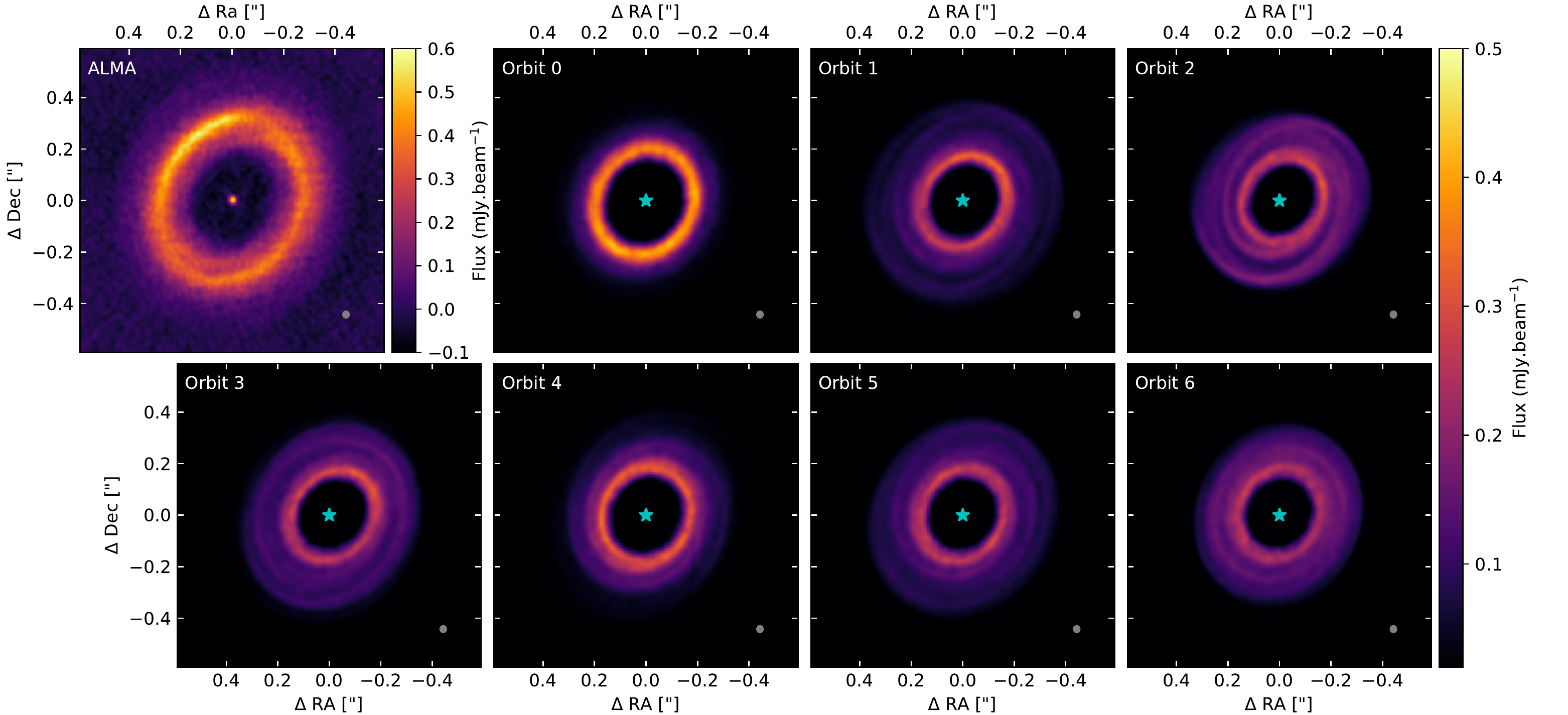}}
\caption{Band~6 dust continuum images. Top left: ALMA observations, then from left to right and top to bottom: synthetic observations for orbits 0 to 6. The $25\times23$~mas beam with PA~=~$2\degr$ is shown in the lower-right corner. The ALMA image shows the unresolved inner disc at the center. In the synthetic images, the position of the central star is marked in cyan. Note the difference in scale with Figs.~\ref{Fig:CO3-2mom0} and \ref{Fig:img0.79}.}
\label{Fig:img1300}
\end{figure*}

\subsubsection{Scattered light}
\label{Sec:ScatteredLight}

Fig~\ref{Fig:img0.79} shows synthetic scattered light images at $\lambda=0.79\ \mu$m ($I'$~band) of the disc for our seven orbital configurations, with the same angular resolution and coronagraph size as in the VLT/SPHERE image from \citet{Benisty2017}, shown in the top left panel. We chose this band rather than the $R'$ or $J$ band because it shows the most contrasted image in their Fig.~1. None of the seven synthetic images reproduces the observed one, as is to be expected since our disc inner radius is smaller that the observed value (see Section~\ref{Sec:SynthDustObs}). In addition, since we do not include the inner disc, we cannot reproduce the two shadows. Nonetheless, a few observations can be made: the images for orbits~3 and 6 show arc-like features rather than clear spirals while for orbit~2, the spirals do not have the correct orientation. For orbits~1, 4 and 5, the spirals start at the correct angular position but appear too tightly wound for orbits~1 and 5, while their opening angle for orbit~4 appears closer to the observations. For orbit 0, a faint spiral can be seen to the southwest close to the bright inner rim, while to the northeast the disc edge has a spiral shape but without any brightening. Both features have a seemingly correct orientation and opening angle, although they are much fainter than the observed ones. This alone is not enough to conclude, but it supports our choice of orbit~0 as the best one based on $^{12}$CO emission. This criterion also allows orbit~4 as a possible match.

Note that our synthetic scattered light images for orbit~6 appear very different from those computed by \citet{Wagner2018}, despite our choice for this orbit of values of $a$, $e$, and $i$ resembling theirs. This is likely because these authors used different (and unspecified) values for $\Omega_\star$ and $\omega$, possibly resulting in a different $i_\mathrm{rel}$ as well (see also Sect.~\ref{Sec:Intro}).

\subsubsection{Millimetre continuum}
\label{Sec:mm_continuum}

Fig~\ref{Fig:img1300} shows synthetic dust continuum images at $\lambda=1.3$~mm of the disc for our seven orbital configurations, with the same angular resolution as in the combined ALMA band~6 datasets shown in the top left panel and described in Section~\ref{Sec:ALMAdata}. Similarly to the scattered light synthetic images, we cannot derive firm constraints from the millimetre continuum images since our disc inner radius is smaller than the observed one. However, here again we find a qualitative agreement between orbit 0, for which we see a narrow and well defined ring, containing two low-contrast spiral arms, and the ALMA data. Orbits~1, 2, 3, 5 and 6 result in a less contrasted ring at the disc inner edge, which has the wrong orientation for orbits 2 and 3, interior to a more extended and fainter emission. Orbit~4 presents an intermediate case, both in terms of inner ring contrast and extension of the fainter emission.

\subsection{Summary}

\begin{table}
\caption{Summary of our comparison to observations}
\label{Table:summary}
\centering
\begin{tabular}{l@{}cccccccc}
\hline
Criterion && \multicolumn{7}{c}{Orbit} \\ \cline{3-9}
&& 0 & 1 & 2 & 3 & 4 & 5 & 6 \\
\hline
\multicolumn{9}{c}{Main evidence: $^{12}$CO gas} \\
\hline
Large-scale disc morphology && \cmark & \xmark     & \xmark & \xmark & \xmark & \xmark & \xmark \\
Spirals                     && \cmark & \xmark     & \xmark & \xmark & \xmark & \xmark & \xmark \\
Velocity structure          && \cmark & \cmark & \xmark & \xmark & \xmark & \xmark & \xmark \\
\hline
\multicolumn{9}{c}{Circumstantial evidence: dust} \\
\hline
$I'$~band scattered light   && \textbf{?} & \xmark & \xmark & \xmark & \textbf{?} & \xmark & \xmark \\
ALMA band 6 dust continuum  && \textbf{?} & \xmark     & \xmark & \xmark & \textbf{?} & \xmark & \xmark \\
\hline
\end{tabular}
\end{table}

Table~\ref{Table:summary} presents a summary of our comparison of synthetic observations for each of the four orbital configurations to the observed data, with our criteria divided in two groups. The first one contains the main evidence based on $^{12}$CO gas data and presented in Section~\ref{Sec:SynthGasObs}: the large-scale disc morphology and the spiral structure from moment 0 maps and the kinematic structure from channel maps. The second group is made of circumstantial evidence from dust observations in scattered light and mm continuum, discussed in Section~\ref{Sec:SynthDustObs}. Orbits~1 and 4 seem compatible with some features but not others, and orbits 2, 3, 5 and 6 are ruled out by all criteria. Unsurprisingly, the less inclined orbits cause a stronger disc truncation. With the constraint of passing through the observed positions of the secondary star, this leads to a disc that is too small to reproduce its observed size. While the dust data do not allow us to conclude for orbit 0, it is compatible will all criteria in the main group. We therefore conclude that the best orbital solution is indeed orbit 0.

Note that during the revision of this paper, we acquired access to an additional, more recent astrometric data point. Updated orbital fits presented in Appendix~\ref{App:NewOrbitalFits} result in very similar probability distributions for all orbital elements except the eccentricity, which no longer peaks at zero but at $\sim0.1$, reaffirming the validity of our choice of orbits for this study. Only orbit~0 has an eccentricity that does not fall in the 68\% confidence interval of the updated fits but that was already the case for the previous ones (see also Fig.~\ref{Fig:MCMC_chi2_e}).

\section{Discussion}
\label{Sec:Discussion}

\subsection{Can the circumprimary disc and the secondary's orbit be misaligned?}
\label{Sec:DiscuMisalign}

\cite{Wagner2018} argued that because of tidal and viscous dissipation within the disc, one would expect the binary orbit to be relatively circular and coplanar with the circumprimary disc, citing previous work by \citet{Papaloizou1995}, \citet{Bate2000} and \citet{Lubow2000}, who studied the tidal perturbation of a circumprimary disc by a companion on an inclined circular orbit. The best fitting orbits derived by \citet{vanderPlas2019} and presented in Section~\ref{Sec:MethSetup} contradict this statement. Additionally, even though the probability distributions of the orbital fits are not totally inconsistent with a close-to-coplanar orbit, the resulting disc morphology does not reproduce the observations, either for a strictly coplanar and circular orbit \citep{vanderPlas2019} or for a weakly-inclined one (orbit~6, this work). \citet{Bate2000} found that the disc should precess with a period of $\sim20$ binary periods and align with the binary orbital plane on a viscous evolution timescale, evaluated at $\sim100$ precession periods, unless hydrodynamic parametric instabilities develop, in which case the alignment timescale would be comparable to the precession period. However, these numerical estimates have been obtained for equal-mass stars and assuming $\cos i_\mathrm{rel}\sim1$. Coming back to their equations, one can write the ratio of the precession period to the binary period as
\begin{equation}
\frac{T_\mathrm{p}}{T_\mathrm{b}}=\frac{1}{K\cos i_\mathrm{rel}} \frac{\sqrt{1+q}}{q}\left(\frac{R_\mathrm{out}}{a}\right)^{-3/2},
\label{Eq:T_prec_circular}
\end{equation}
where $q=M_\mathrm{B}/M_\mathrm{A}$ is the mass ratio and
\begin{equation}
K=\frac{3}{4}R_\mathrm{out}^{-3/2}\frac{\displaystyle\int_{R_\mathrm{in}}^{R_\mathrm{out}}\Sigma(r) r^3\,\mathrm{d}r}{\displaystyle\int_{R_\mathrm{in}}^{R_\mathrm{out}}\Sigma(r) r^{3/2}\,\mathrm{d}r}
\label{Eq:K_Terquem}
\end{equation}
\citep{Terquem1998}. Its expression for a power-law disk with $\Sigma\propto R^{-p}$ is
\begin{equation}
K=\frac{3}{8}\frac{5-2p}{4-p}\frac{1-(R_\mathrm{in}/R_\mathrm{out})^{4-p}}{1-(R_\mathrm{in}/R_\mathrm{out})^{5/2-p}}.
\label{Eq:K_power_law}
\end{equation}
For our parameters $p=1$, $R_\mathrm{in}=12$~au and $R_\mathrm{out}=60$~au, this gives $K=0.41$, while if $R_\mathrm{in}=0$, $K=3/8=0.375$ for all $R_\mathrm{out}$. For circular orbits of radius $a\sim100$ or 200~au and $i_\mathrm{rel}\sim60\degr$, one finds $T_\mathrm{p}\sim100$ or 270~$T_\mathrm{b}$. Within a factor of $\sim2$, this is compatible with the precession periods we observe in our simulations (see Section~\ref{Sec:Forward}). \citet{Bate2000} suggest that, to a first approximation, the time-averaged potential of the secondary can be obtained by replacing in the expression of $T_\mathrm{p}$ the radius of a circular orbit by $a\sqrt{1-e^2}$ for an eccentric orbit, which modifies equation~(\ref{Eq:T_prec_circular}) as follows
\begin{equation}
\frac{T_\mathrm{p}}{T_\mathrm{b}}=\frac{1}{K\cos i_\mathrm{rel}} \frac{\sqrt{1+q}}{q}\left(\frac{R_\mathrm{out}}{a(1-e^2)}\right)^{-3/2}.
\label{Eq:T_prec_eccentric}
\end{equation}
This gives a value of $T_\mathrm{p}\sim250\ T_\mathrm{b}$ for orbit 0. 

The expression given by \citet{Bate2000} for the alignment timescale can be written for an eccentric orbit as
\begin{equation}
\frac{t_\mathrm{align}}{T_\mathrm{p}}\sim\frac{1}{K\cos i_\mathrm{rel}\,q\,\alpha_\mathrm{SS}} \left(\frac{H}{R}\right)^2\left(\frac{R_\mathrm{out}}{a\sqrt{1-e^2}}\right)^{-3},
\label{Eq:T_align_eccentric}
\end{equation}
which, with our values of $\alpha_\mathrm{SS}=5\times10^{-3}$ and $H/R=0.05$, gives $t_\mathrm{align}\sim750\ T_\mathrm{p}\sim187,500\ T_\mathrm{b}$ for orbit 0. With $T_\mathrm{b}=2161$~yr, this amounts to $\sim400$~Myr, much longer than disk lifetimes. In the worst-case scenario where strong parametric instabilities do develop, the alignment timescale is given by
\begin{equation}
\frac{t_\mathrm{align}}{T_\mathrm{p}}\sim\frac{1}{(K\cos i_\mathrm{rel})^2\sin i_\mathrm{rel} \,q^2} \left(\frac{H}{R}\right)^4\left(\frac{R_\mathrm{out}}{a\sqrt{1-e^2}}\right)^{-6}.
\label{Eq:T_align_param_inst}
\end{equation}
For orbit 0, $t_\mathrm{align}\sim16\ T_\mathrm{p}\sim4000\ T_\mathrm{b}\sim8.6$~Myr. \citet{Bate2000} caution that this estimate is probably a lower limit, and the effect is likely to be even weaker for an eccentric orbit. It seems therefore possible for inclined discs to maintain a substantial misalignment with the orbit of an eccentric external companion for most of their lifetime.

Indeed, this is in line with recent observations of $\sim1$--2 Myr-old transition discs exhibiting shadows or dips in scattered light. These illumination features are thought to be caused by a misaligned inner disc that obstructs the stellar light \citep{Marino2015,Min2017}. The detailed modelling of individual sources, such as HD~142527 \citep{Price2018} and AB~Aur \citep{Poblete2020}, shows how a circumprimary disc can form from the infalling material in the presence of an eccentric and inclined companion inside the disc cavity.

Note however that the circumbinary ring that forms in our simulations from gas initially in the outer disc and flung out seems coplanar with the binary orbit (as best seen for orbits~2, 3, 5 and 6 in Fig.~\ref{Fig:Rewind10}), likely because of angular momentum exchange with the secondary star.

On a related note, the system's age was previously estimated by \citet{Collins2009} to $10\pm2$~Myr, by comparing the secondary's spectral type to pre-main sequence H-R diagrams. More recently, \citet{Vioque2018} derived an age of $6.5\pm0.5$~Myr for the primary by fitting isochrones to Gaia data. This younger age is more in line with typical disc lifetimes as well as compatible with the most pessimistic alignment timescale. If the secondary's age determination is correct, there remains the possibility that both stars did not form together and that the secondary was later captured. This would provide an explanation for its inclined, eccentric orbit. In that case, its interaction with the disc may have lasted for much less than a few Myr, an argument that would also be in favour of the non-coplanarity of the orbit and the disc (this is also discussed in \citetalias{PaperII}).

Determining the formation pathway of any binary system is difficult because both internal and external dynamical processes alter the population of multiple star systems from their birth onwards \citep{Reipurth2014}. In their early phases, substantial accretion of gas is able to change the orientation of disc planes and no mechanism -- cloud or disc fragmentation, star-disc encounters, accretion, dynamical evolution -- dominates in creating the variety of multiple systems \citep{Bate2018}. The capture hypothesis suggested by the age difference would still make HD~100453 a rather rare, and thus very interesting, system.

\subsection{Origin of the spirals}
\label{Sec:DiscuSpirals}

Spiral density waves are a natural outcome of the gravitational interaction between a disc and a companion \citep[e.g.][]{Rafikov2002}. In previous work, \citet{Dong2016,Wagner2018,Rosotti2020} performed numerical simulations of HD~100453's system with the secondary on a coplanar, circular orbit and all found that the companion launched spirals resembling the observed ones. In our simulations, the companion on an inclined, eccentric orbit causes spiral arms of different strengths, which we argue are a better match to the observations of $^{12}$CO gas.

As pointed out by \citet{Rosotti2020}, the hypothesis that the spirals are caused by the shadows cast by the inner disc is unlikely. Indeed, the fact that the southern spiral points to the secondary would be an exceptional coincidence. One might argue that it is an equally exceptional coincidence that the shadows are at the base of the observed spirals in scattered light, suggesting a causal relationship. However, \citet{Cuello2019} showed that this mechanism would not produce observable sub-mm continuum spirals, such as those seen in the top left panel of Fig.~\ref{Fig:img1300}. We therefore conclude that the origin of the grand design spiral structure observed in HD~100453's disc is the gravitational interaction with the secondary star.

\subsection{Caveats}

The main caveat of this work is the small number of orbital configurations tested and the fact that the best solutions coming out of the MCMC astrometric fit are not unique. It would have been impractical and extremely time-consuming to perform SPH simulations for dozens of orbital solutions. However, our selection of 7 orbits is reasonably representative of the orbital element probability distributions among the best fits and, while the first three have rather similar values of the reduced $\chi^2$, the other four already has a significantly larger value. We thus feel confident that the true orbit of the binary system is similar to orbit 0, and that in particular it is inclined with respect to the disc plane by an angle close to $60\degr$ and significantly eccentric.

One may argue that our methodology of visually comparing synthetic images or channel maps to observations is somewhat subjective compared to a more objective astrometric fitting. However, the very small observed orbital arc leads to probability distributions allowing a wide range of orbital parameters, many of them providing very good fits to the astrometric data, around the most probable ones (see also Appendix~\ref{App:NewOrbitalFits}). The whole purpose of this work is to provide independent constraints to narrow down the parameter space. For example, a semi-major axis of $\sim100$~au is favoured by orbital fits but our simulations of such orbits result in a tension with the observations. If the companion is instead on an orbit similar to orbit 0, most of the observed disc structures are recovered. One has to keep in mind that the most probable orbit is not necessarily the true one, especially when it causes a disc morphology that is incompatible with observations.

Running numerical simulations requires the choice of input parameters and initial conditions, which can be a source of uncertainty. We use the available observations to constrain the disc mass and size as much as possible. We choose reasonable values for other disc parameters such as the initial dust-to-gas ratio, viscosity, aspect ratio or surface density and temperature profile exponents. The comparison of our forward and rewind simulations shows that the initial tilt and twist angles only affect the final disc orientation but result in almost indistinguishable disc morphologies. An extensive parameter sweep of all input variables is in any case out of the scope of this work.

The outer disc in our synthetic images is brighter than in the observations, this is particularly visible in the $^{12}$CO channel maps (Fig.~\ref{Fig:CO3-2cm_obs_orb0}). This brightness discrepancy is independent of the orbit choice and likely due to the fact that our simulations do not include the inner disc very close to the star. This inner disc can intercept a fraction of the stellar photons, reducing the illumination of the outer disc. The synthetic observations computed in \citetalias{PaperII} do include the inner disc and indeed show brightness levels closer to the observed ones. In this paper, we do not seek absolute calibration. We focus on the shape and location of structures in the velocity channel maps, which depend on the interaction with the secondary and not on the absolute value of the brightnesss temperature. Therefore, this does not affect our conclusions.

We use a locally isothermal equation of state, and as a consequence cannot recover the effects of a vertical temperature gradient on the opening angles of the spiral arms, as evidenced by \citet{Rosotti2020}. In particular, the different tracers we consider in this paper -- CO gas lines, mm dust continuum, optical scattered light -- probe different altitudes in the disc and would show different opening angles. \citet{Rosotti2020} found a $\sim10\degr$ difference between the extremes in the mid-plane or at the disc surface. Given the differences between orbits in the moment 0 maps, a $10\degr$ uncertainty would likely have a small impact on our conclusions.

The formation of a circumsecondary disc is another expected outcome from the gravitational interaction between the disc and the secondary star \citep[e.g.][]{Menard2020}. In the coplanar case \citep{vanderPlas2019}, such a disc indeed forms. In this study, a transient one forms but does not survive long for the most inclined orbits (see Video 2). However, we must point out that we use a sink radius for the secondary star that is much larger than the physical accretion radius, resulting in an artificially faster accretion \citep[the same effect was seen by][for the circumprimary disc of HD~142527]{Price2018}. Therefore a circumsecondary disc might very well survive over secular timescales.

\section{Conclusions}
\label{Sec:Concl}

In this paper we revisit the orbit of the binary star system HD~100453, previously assumed to be circular and coplanar with the disc. Astrometric fits performed by \citet{Wagner2018} and \citet{vanderPlas2019} are poorly constrained because the available measurements only span a very small arc of the orbit. We seek to further constrain the orbit via the observation of the disc structures caused by the gravitational interaction with the secondary star.

We select seven orbits among the best orbital fits and perform numerical simulations of the gas and dust disc evolution for each configuration. We then compute synthetic observations from the resulting gas and dust distributions and compare them with ALMA observations of the $^{12}$CO gas emission. While not as constraining, scattered light and mm continuum observations support our conclusions.

Our findings are the following:
\begin{itemize}
    \item Orbit 0, despite being different from the most likely orbit (based on astrometric fitting), is the only solution within our sample that simultaneously reproduces most of the observed disc features. It is inclined with respect to the disc by $61\degr$, with an eccentricity $e=0.32$ and a semi-major axis $a=207$~au, twice as large as previously assumed. Although the solution is certainly not unique, the true orbit likely has a similar inclination and is significantly eccentric.
    \item Orbit~1, almost circular but still highly inclined ($58\degr$) with respect to the disc plane, is not a very good match, but it does reproduce a few of the observed features.
    \item Orbit~4, with intermediate values of its semi-major axis (141~au) and relative inclination ($36\degr$) seems a possible match for the dust observations, but fails to reproduce the spirals or kinematics of the CO gas.
    \item All other considered orbits, with lower values of their semi-major axis, eccentricity and/or relative inclination, are inconsistent with the observational data.
    \item The disc precesses around the orbit's angular momentum vector but the 
    relative inclination between the orbit and disc planes evolves slowly. This is compatible with alignment timescale for unequal stellar masses and eccentric orbits.
    \item Finally, we confirm that the well-defined spiral structure observed in the disc is caused by the gravitational interaction with the secondary star.
\end{itemize}
HD~100453 constitutes a wonderful laboratory for disc dynamics in binary systems. The misalignment and the eccentricity of the binary orbit are key to understanding (circumstellar) disc structure and evolution. \citetalias{PaperII} explores the dynamical effects of a hidden companion within the central disc cavity.

\section*{Acknowledgements}
The authors thank the anonymous referee for their prompt and constructive report.
We thank Myriam Benisty and Maria Giulia Ubeira-Gabellini for giving us access to their $I'$-band data from \citet{Benisty2017} and VLT/SPHERE-IRDIS 2019 data, respectively. The authors acknowledge funding from ANR (Agence Nationale de la Recherche) of France under contract number ANR-16-CE31-0013 (Planet-Forming-Disks) and thank the LABEX Lyon Institute of Origins (ANR-10-LABX-0066) of the Universit\'e de Lyon for its financial support within the programme `Investissements d'Avenir' (ANR-11-IDEX-0007) of the French government operated by the ANR. This project has received funding from the European Union's Horizon 2020 research and innovation programme under the Marie Sk\l{}odowska-Curie grant agreements No 210021 and No 823823 (DUSTBUSTERS). RN acknowledges funding from the European Research Council (ERC) under the European Union's Horizon 2020 research and innovation programme (grant agreement No 681601). A.L.M. acknowledges the financial support of the F.R.S.-FNRS through a postdoctoral researcher grant. C.P. acknowledges funding from the Australian Research Council via FT170100040 and DP180104235. SPH simulations were run on OzStar, funded by Swinburne University of Technology and the Australian government. Figures were made with \textsc{Splash} \citep{Splash2007,SplashCode} and the Python library \texttt{matplotlib} \citep{matplotlib}. This work has made use of the SPHERE Data Centre, jointly operated by OSUG/IPAG (Grenoble), PYTHEAS/LAM/CeSAM (Marseille), OCA/Lagrange (Nice), Observatoire de Paris/LESIA (Paris), and Observatoire de Lyon (OSUL/CRAL).

\section*{Data Availability}
The {\sc Phantom} SPH code is available from \url{https://github.com/danieljprice/phantom}. {\sc mcfost} is available for use on a collaborative basis from \url{https://ipag.osug.fr/~pintec/mcfost/docs/html/overview.html}. The input files for generating our SPH simulations and radiative transfer models, as well as the fitted orbits from the OFTI and MCMC algorithms, are available on request.



\bibliographystyle{mnras}
\bibliography{orbit_HD100453}




\appendix
\section{Additional figures}
\label{App:AdditionalFigs}

\subsection{Gas+dust simulations}
\label{App:FigGDsims}

\begin{figure}
\centering
\resizebox{\hsize}{!}{
\includegraphics{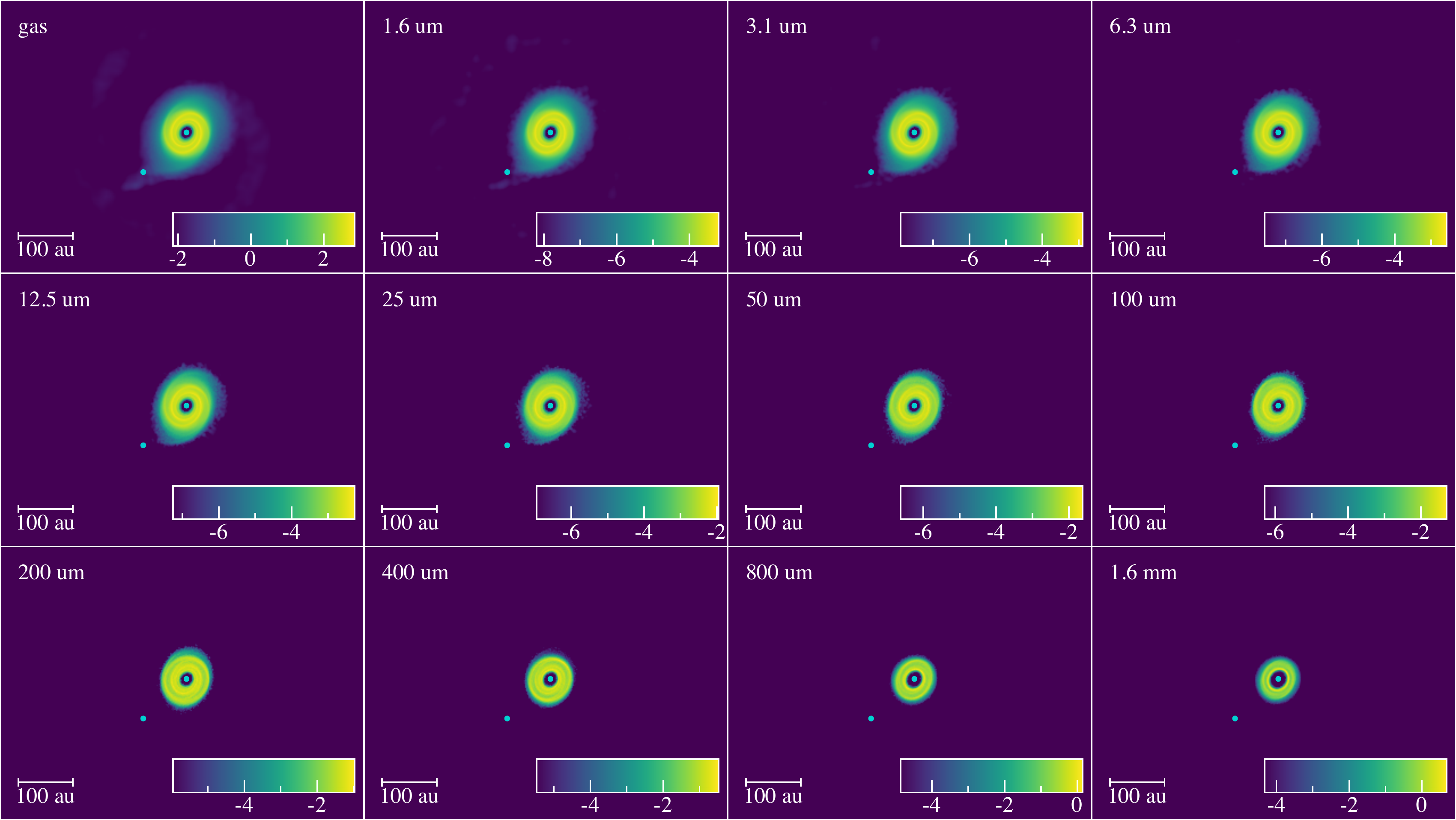}
}
\caption{Same as Fig.~\ref{Fig:GDorb0} for orbit 1.}
\label{Fig:GDorb1}
\end{figure}

\begin{figure}
\centering
\resizebox{\hsize}{!}{
\includegraphics{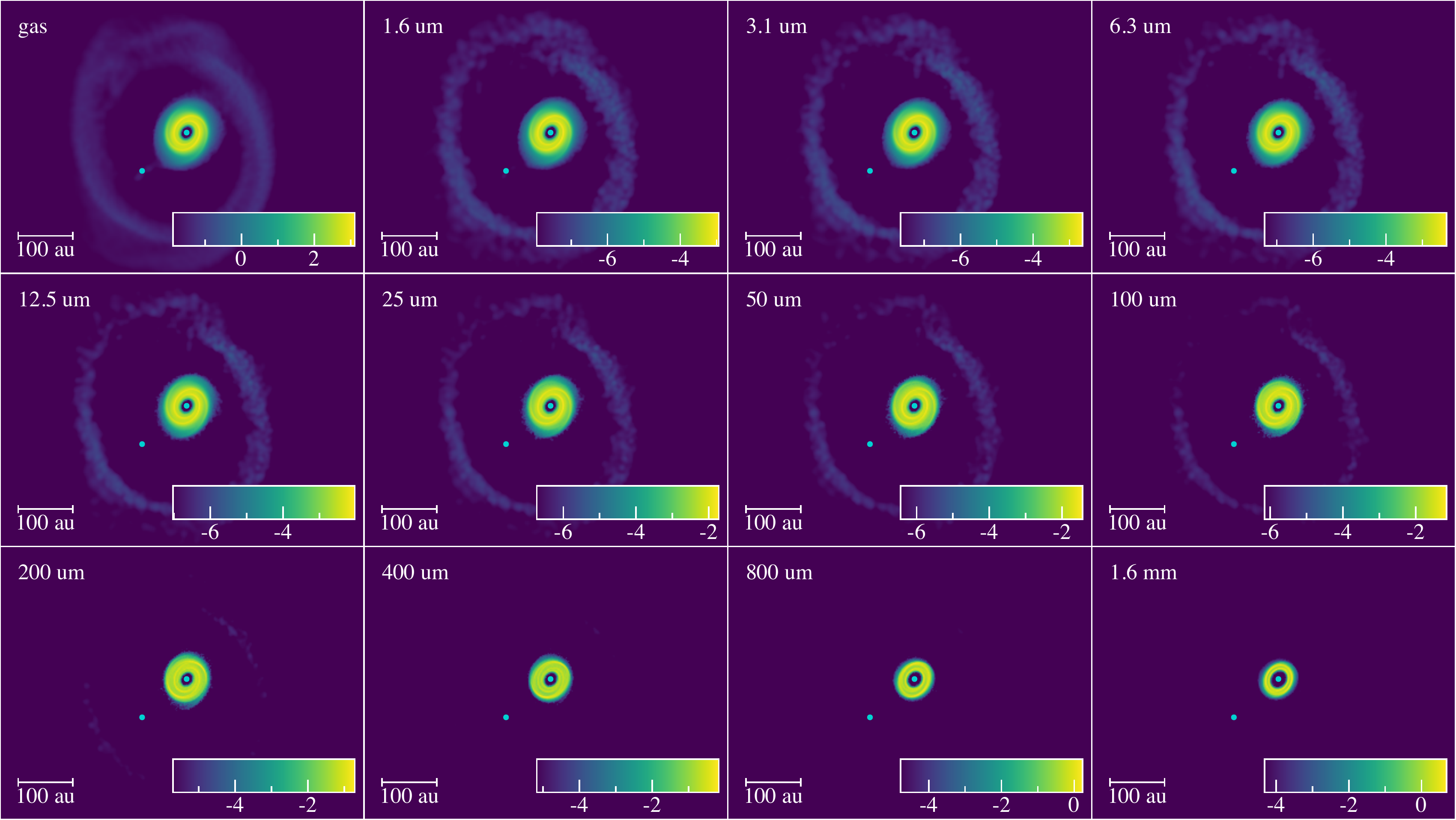}
}
\caption{Same as Fig.~\ref{Fig:GDorb0} for orbit 2.}
\label{Fig:GDorb2}
\end{figure}

\begin{figure}
\centering
\resizebox{\hsize}{!}{
\includegraphics{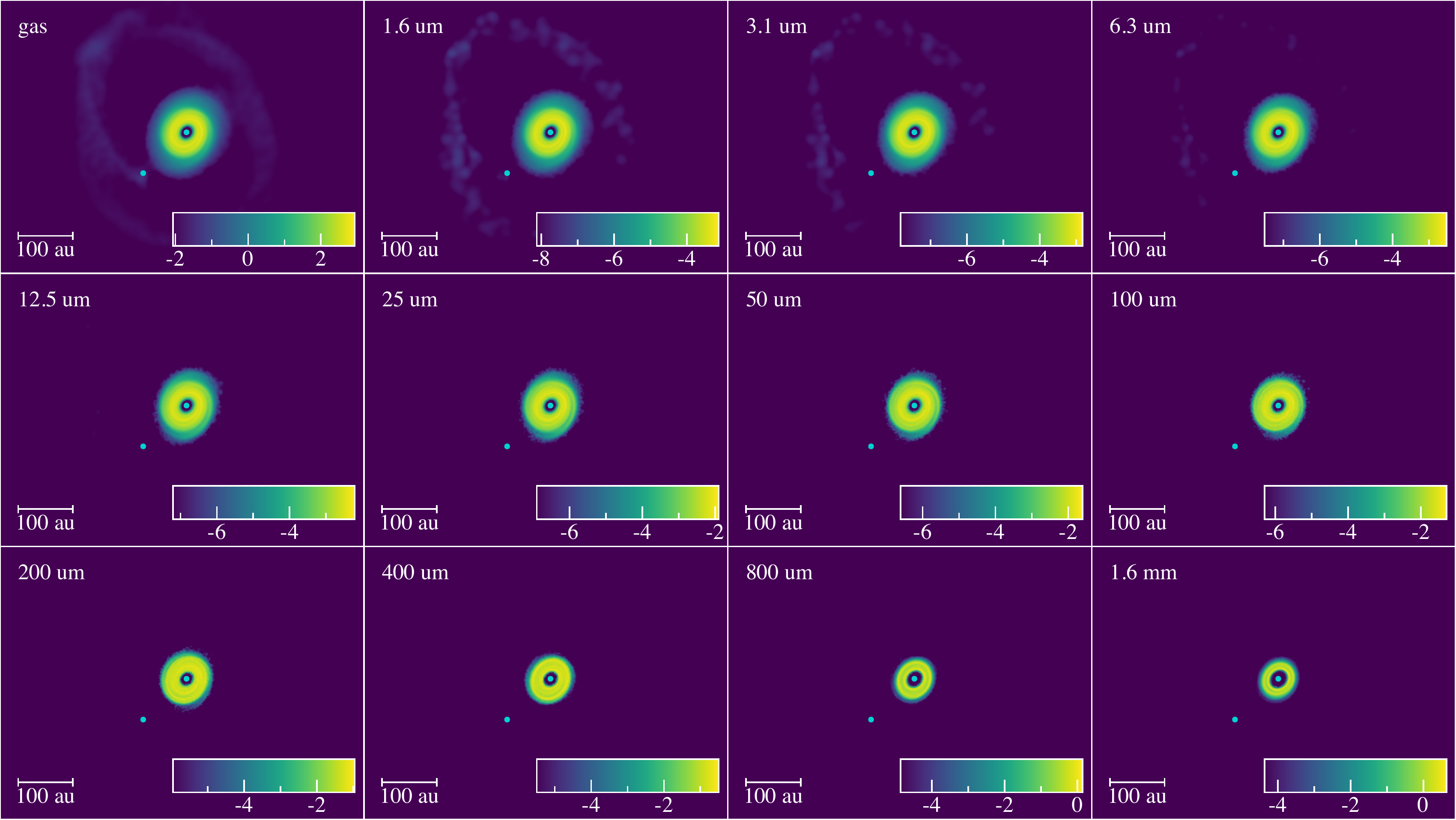}
}
\caption{Same as Fig.~\ref{Fig:GDorb0} for orbit 3.}
\label{Fig:GDorb3}
\end{figure}

Figures~\ref{Fig:GDorb1}, \ref{Fig:GDorb2} and \ref{Fig:GDorb3} present rendered column density maps of the gas and each of the 11 grain sizes in the plane of the sky at the end of the simulations for orbits~1, 2 and 3, respectively, similarly to Fig.~\ref{Fig:GDorb0} for orbit 0.

\begin{figure}
\centering
\resizebox{\hsize}{!}{
\includegraphics{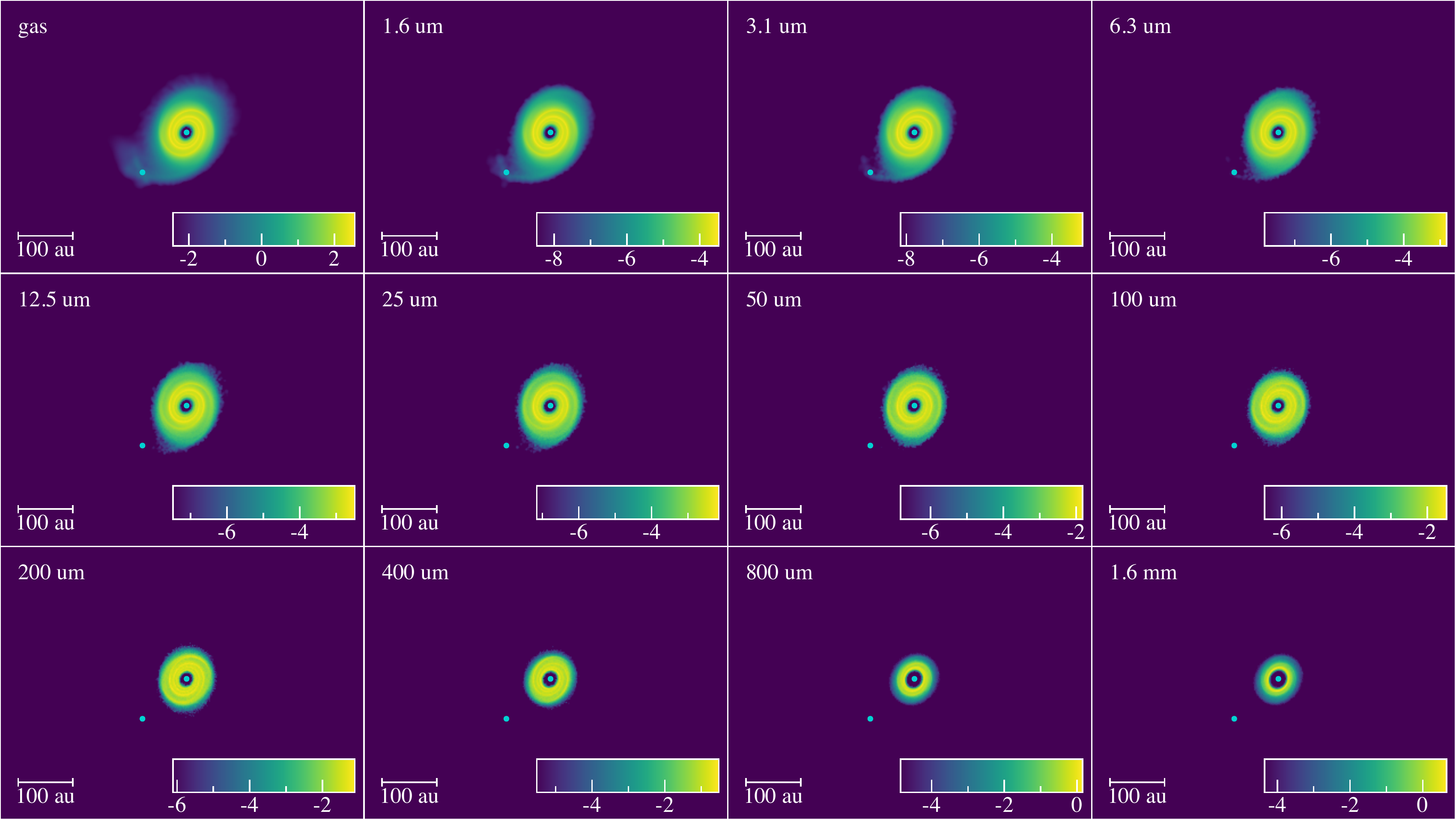}
}
\caption{Same as Fig.~\ref{Fig:GDorb0} for orbit 4. }
\label{Fig:GDorb4}
\end{figure}

\begin{figure}
\centering
\resizebox{\hsize}{!}{
\includegraphics{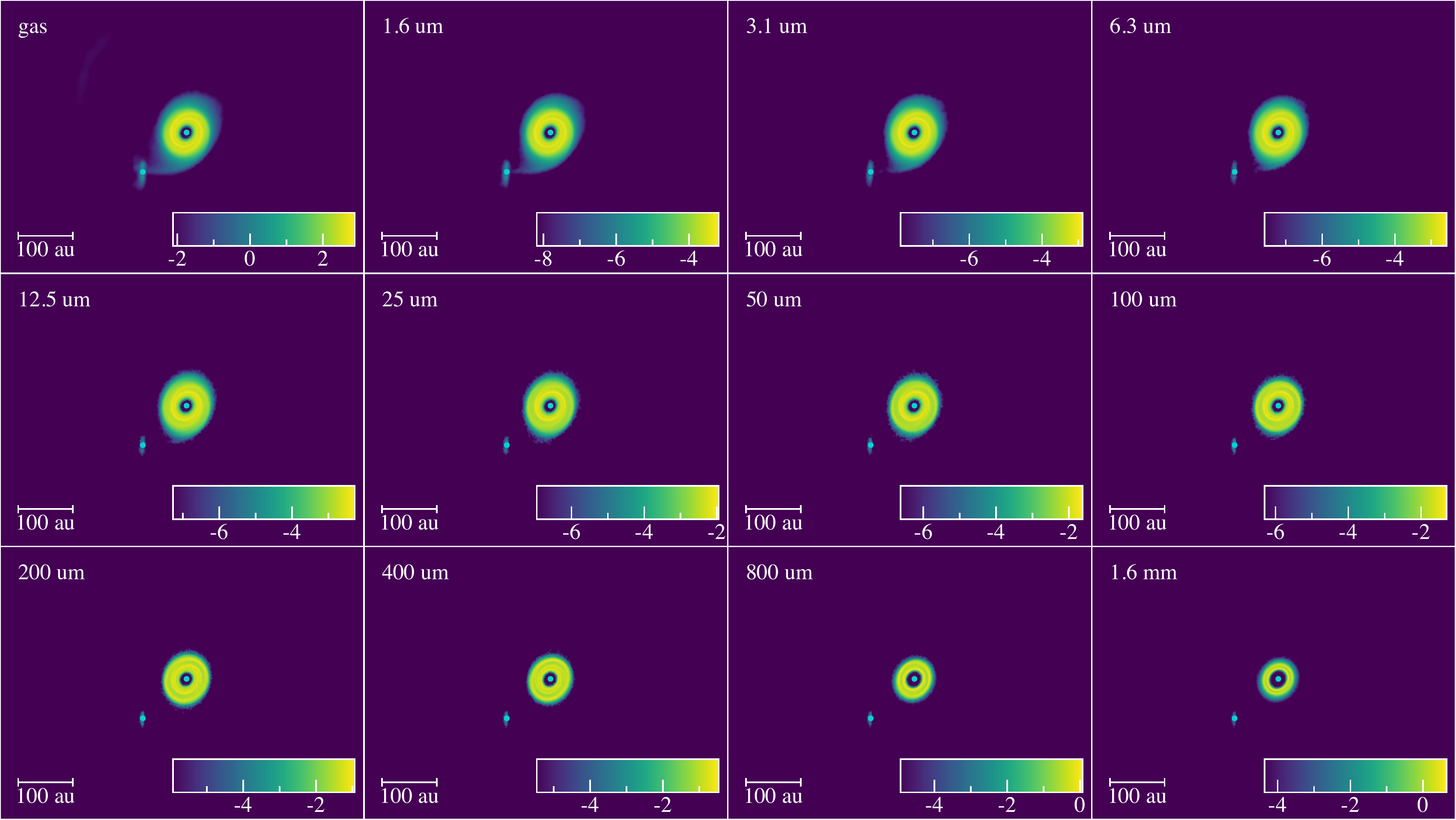}
}
\caption{Same as Fig.~\ref{Fig:GDorb0} for orbit 5.}
\label{Fig:GDorb5}
\end{figure}

\begin{figure}
\centering
\resizebox{\hsize}{!}{
\includegraphics{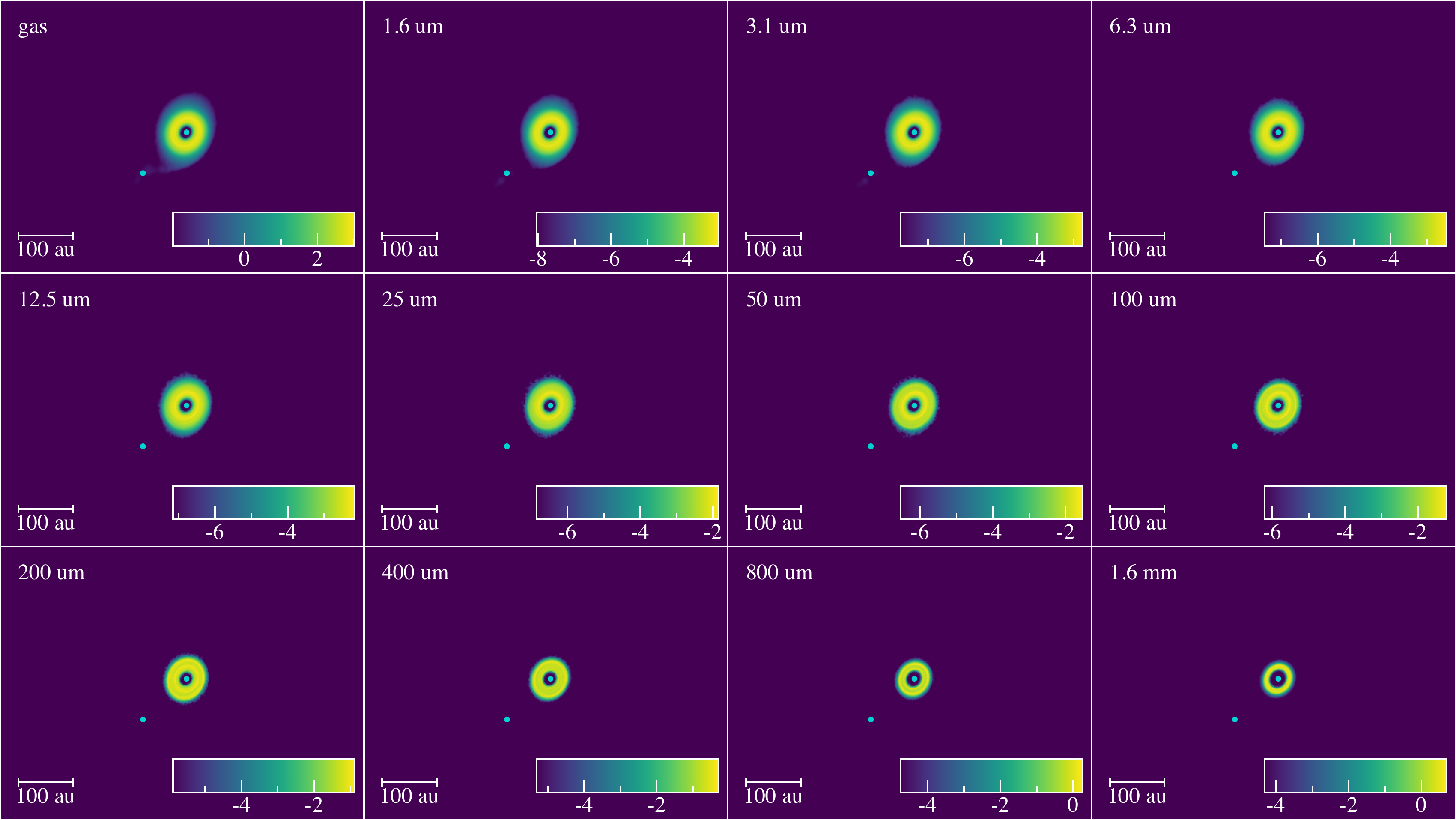}
}
\caption{Same as Fig.~\ref{Fig:GDorb0} for orbit 6.}
\label{Fig:GDorb6}
\end{figure}

\subsection{Synthetic channel maps}
\label{App:FigChannelMaps}

\begin{figure}
\centering
\resizebox{\hsize}{!}{
\includegraphics{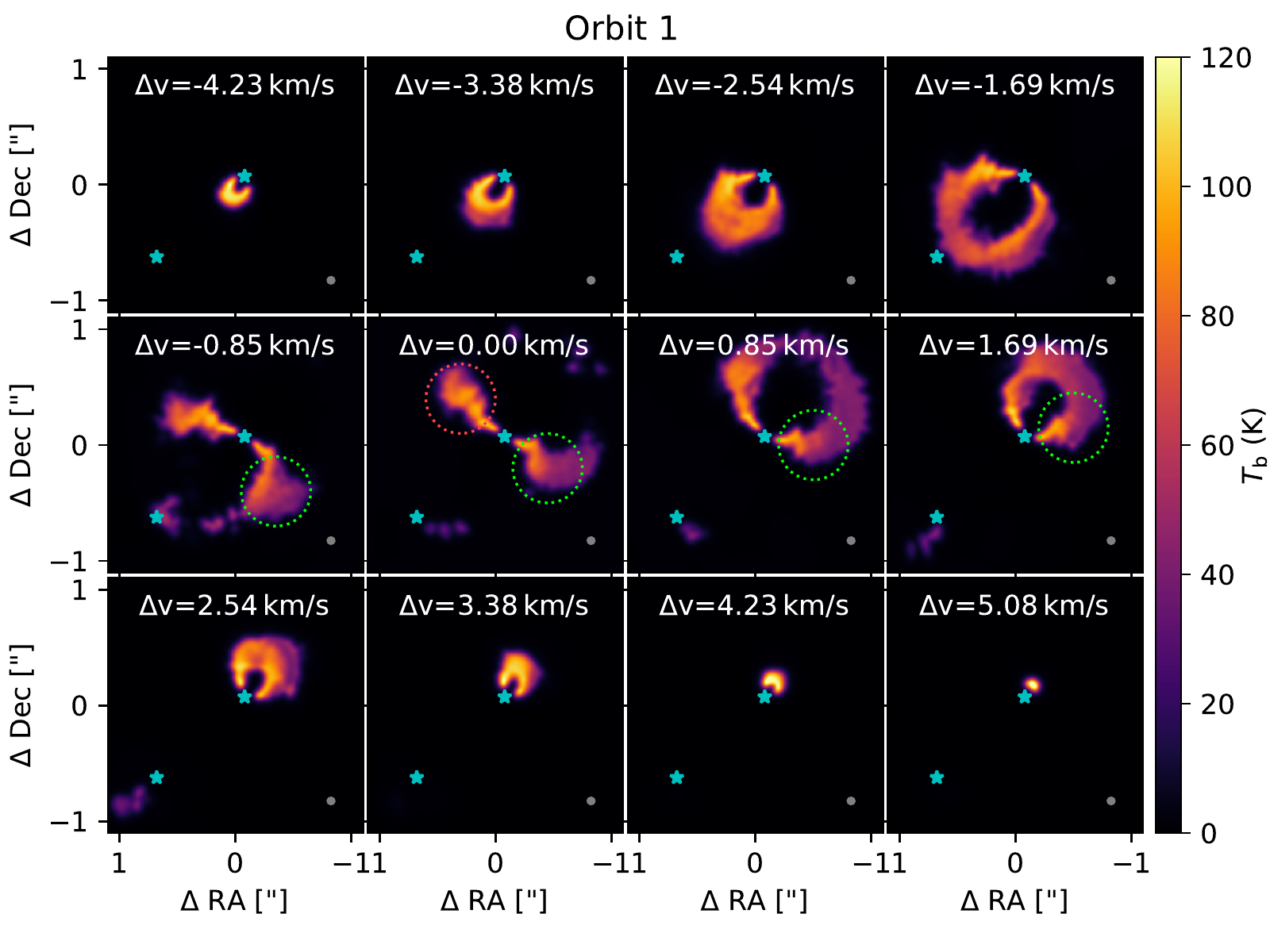}
}
\caption{Synthetic channel maps of the $^{12}$CO $J=3$--2 line for orbit~1. The positions of both stars are marked in cyan and the beam is in the lower-right corner. Dotted circles show the locations of the velocity kinks caused by the spirals in different channels.}
\label{Fig:CO3-2cm_orb1}
\end{figure}

\begin{figure}
\centering
\resizebox{\hsize}{!}{
\includegraphics{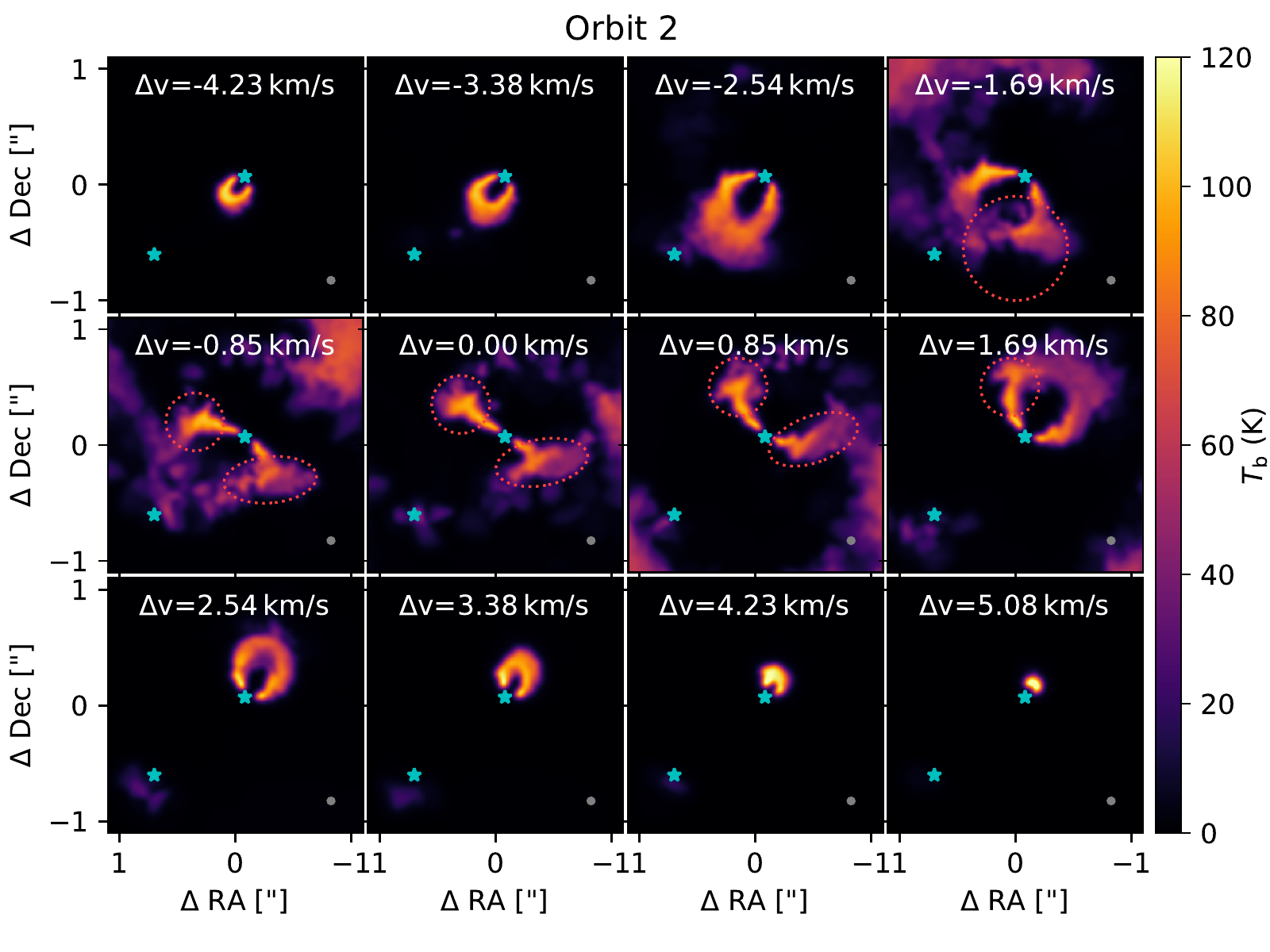}
}
\caption{Same as Fig.~\ref{Fig:CO3-2cm_orb1} for orbit 2.}
\label{Fig:CO3-2cm_orb2}
\end{figure}

\begin{figure}
\centering
\resizebox{\hsize}{!}{
\includegraphics{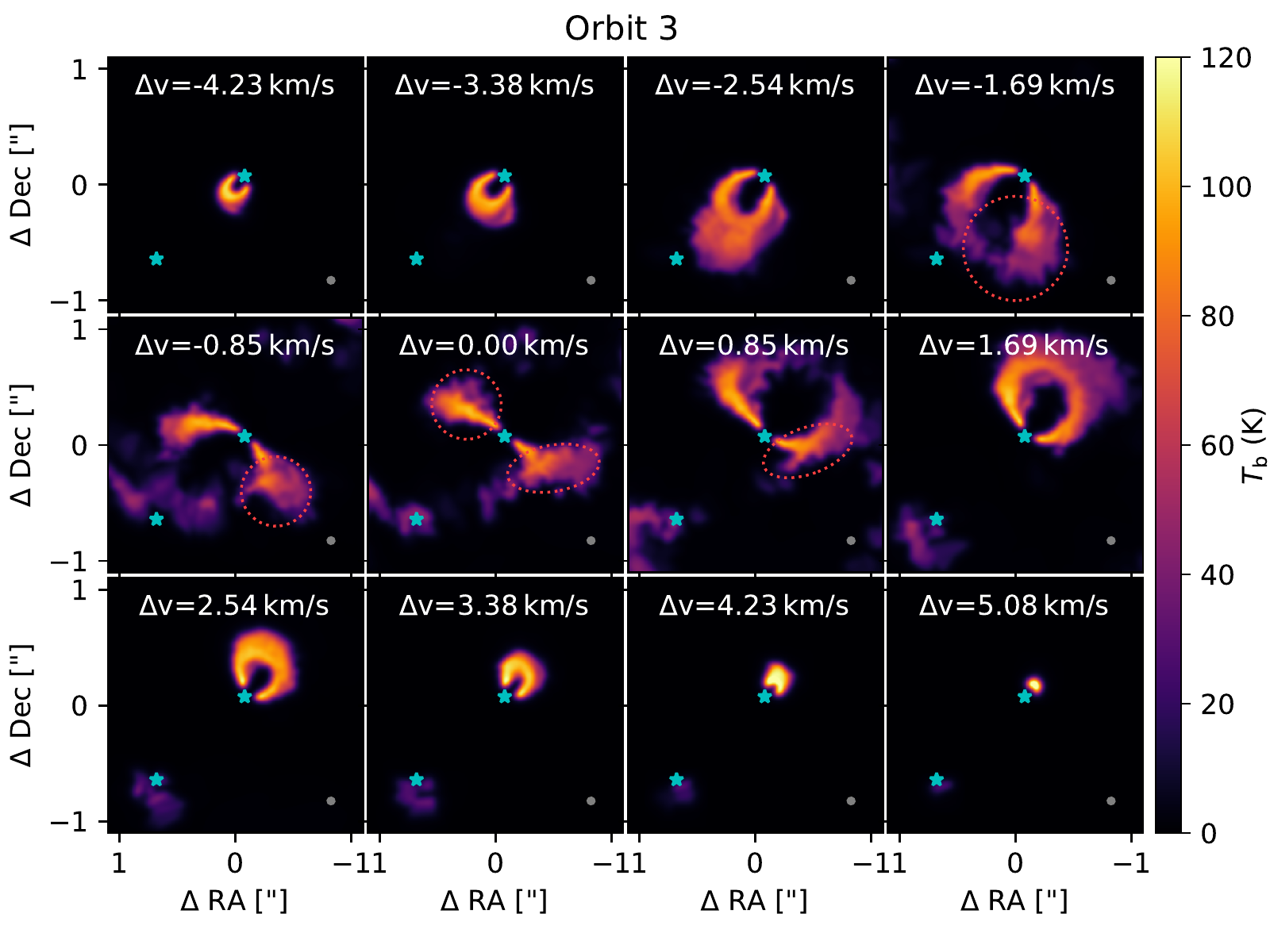}
}
\caption{Same as Fig.~\ref{Fig:CO3-2cm_orb1} for orbit 3.}
\label{Fig:CO3-2cm_orb3}
\end{figure}

\begin{figure}
\centering
\resizebox{\hsize}{!}{
\includegraphics{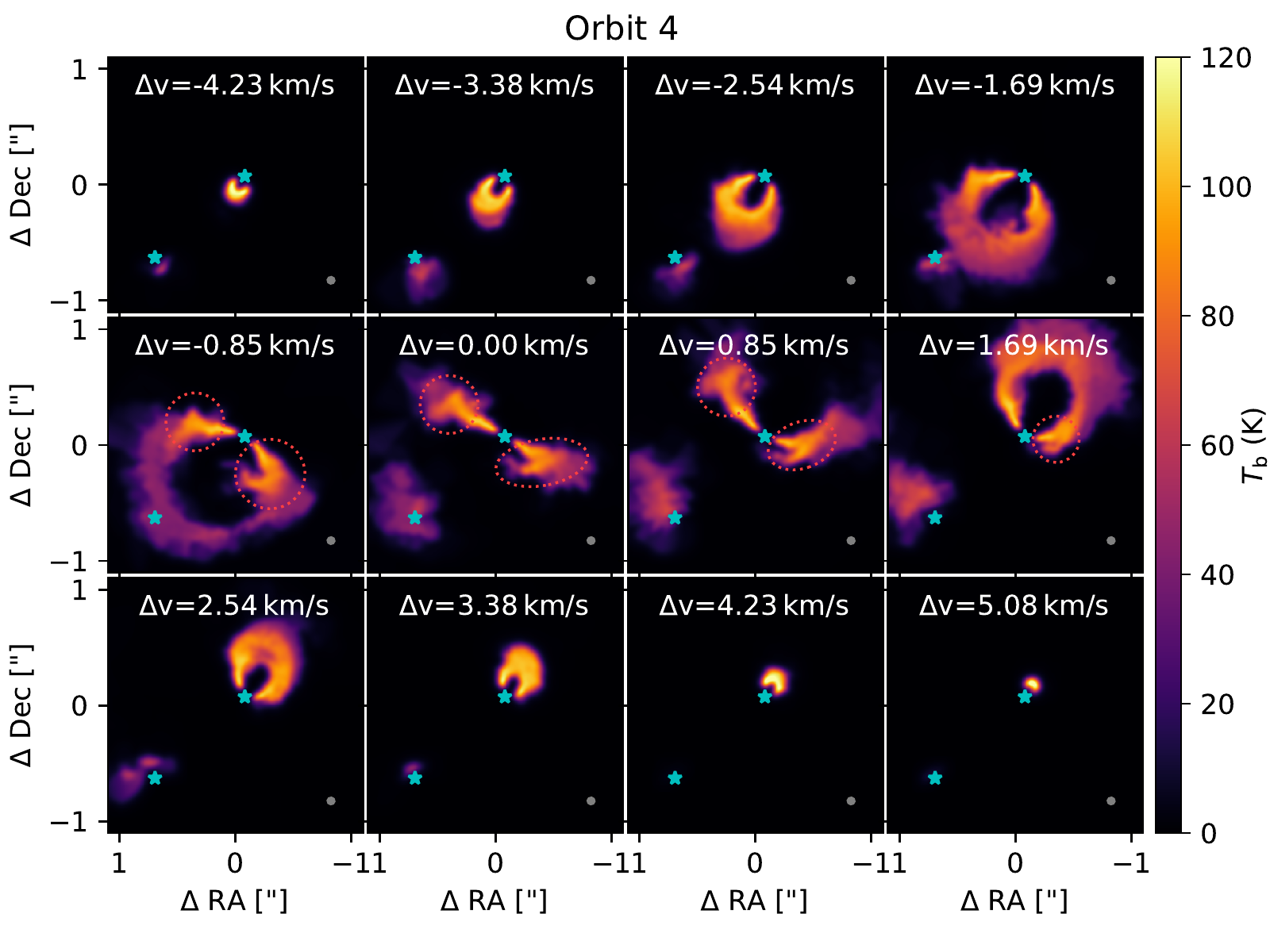}
}
\caption{Same as Fig.~\ref{Fig:CO3-2cm_orb1} for orbit 4.}
\label{Fig:CO3-2cm_orb4}
\end{figure}

\begin{figure}
\centering
\resizebox{\hsize}{!}{
\includegraphics{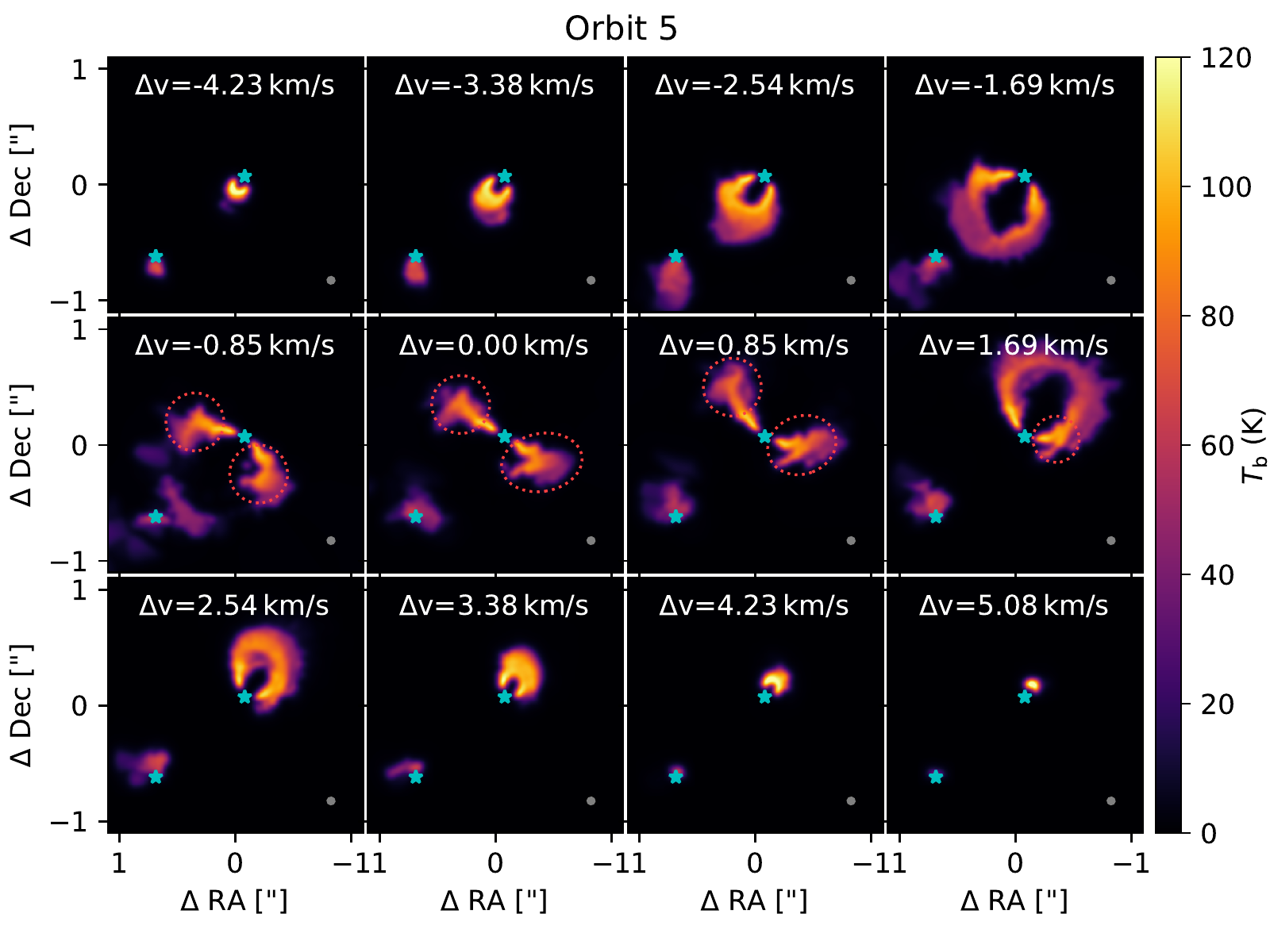}
}
\caption{Same as Fig.~\ref{Fig:CO3-2cm_orb1} for orbit 5.}
\label{Fig:CO3-2cm_orb5}
\end{figure}

\begin{figure}
\centering
\resizebox{\hsize}{!}{
\includegraphics{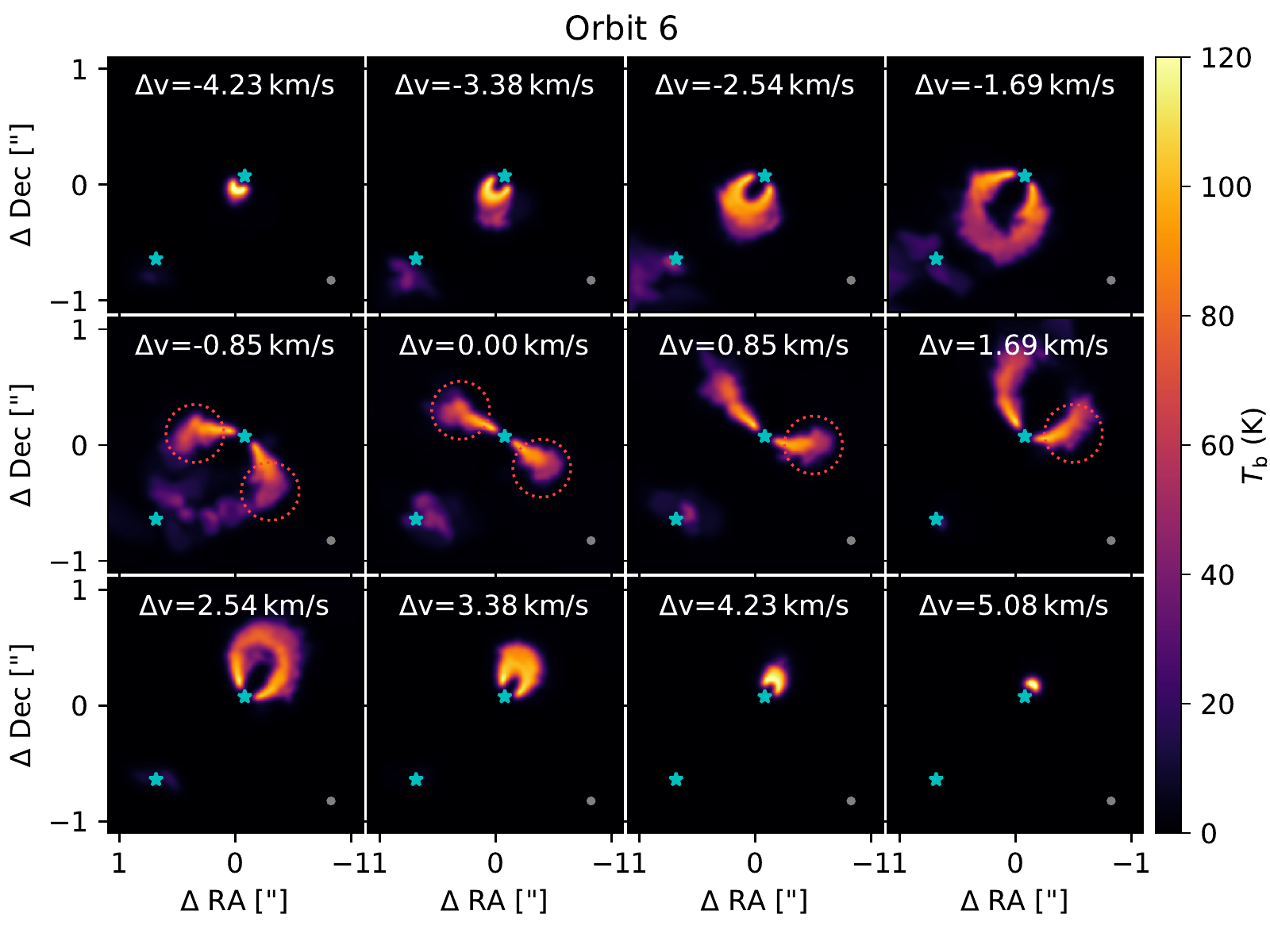}
}
\caption{Same as Fig.~\ref{Fig:CO3-2cm_orb1} for orbit 6.}
\label{Fig:CO3-2cm_orb6}
\end{figure}

Figures~\ref{Fig:CO3-2cm_orb1} to \ref{Fig:CO3-2cm_orb6} show synthetic channel maps of the $^{12}$CO $J=3$--2 line for orbits~1 to 6, respectively, similarly to the right panel of Fig.~\ref{Fig:CO3-2cm_obs_orb0} for orbit 0.

\section{Updated orbital fitting}
\label{App:NewOrbitalFits}

During the revision of this paper, we acquired access to VLT/SPHERE-IRDIS observations of the HD~100453 system from 2019 April 6--7 (Programme 0103.C-0847, PI M.~G. Ubeira-Gabellini), which were reduced with the SPHERE Data Center \citep{Delorme2017} using the public astrometric calibration \citep{Maire2016} and from which we obtained a new astrometric data point complementing the catalogue from \citet{Wagner2018}. The data reduction steps are as follows: We first pre-reduce the IRDIS images to generate good-quality cleaned and re-centered images within a single master cube associated with their parallactic angle values. Subsequent steps aim at the estimation and subtraction of the stellar halo for each image, followed by derotation and stacking of the residuals. The most critical step is estimating the stellar halo that drives the level of the residuals. We apply different Angular Differential Imaging (ADI) algorithms to optimise the detection performances and to identify associated biases. We rely mainly on the SpeCal software \citep{Galicher2018} which offers various ADI options out of which we select the Classical ADI and TLOCI algorithms, which lead to consistent astrometric measurements. After building a model of the point source using the technique from \citet{Galicher2018}, the flux and position of this synthetic image are adjusted to best fit the real point source image that includes the positive and the negative parts of the image for the TLOCI option. In order to have as homogeneous a dataset as possible for a new fit of the binary orbit, we also re-reduced the 2016 SPHERE data in the same way. The astrometry from the 2016 datasets in \citet{Wagner2018} were not consistent with the 2019 astrometry but this new reduction solved the discrepancy. Small systematic errors between \citet{Wagner2018} and our study could arise from differences in the data analysis and/or calibration. The other astrometric point from 2015 was consistent with the new astrometric point and remains unchanged. The updated astrometric data are listed in Table~\ref{Table:NewAstrometry}.

\begin{table}
\caption{Updated astrometric data}
\label{Table:NewAstrometry}
\centering
\begin{tabular}{@{}lllll@{}}
\hline
Date & Instrument & Separation & PA & Ref. \\
\hline
2003 Jun 02 & VLT/NACO   & $1\farcs049\pm0\farcs007$ & $127\fdg2\pm0\fdg3$ & $^\mathrm{a}$ \\
2006 Jun 22 & VLT/NACO   & $1\farcs042\pm0\farcs005$ & $128\fdg3\pm0\fdg3$ & $^\mathrm{a}$ \\
2015 Apr 10 & VLT/SPHERE & $1\farcs047\pm0\farcs003$ & $131\fdg6\pm0\fdg2$ & $^\mathrm{b}$ \\
2016 Jan 16 & VLT/SPHERE & $1\farcs045\pm0\farcs003$ & $132\fdg0\pm0\fdg2$ & $^\mathrm{c}$ \\
2016 Jan 21 & VLT/SPHERE & $1\farcs049\pm0\farcs003$ & $132\fdg1\pm0\fdg2$ & $^\mathrm{c}$ \\
2016 Jan 23 & VLT/SPHERE & $1\farcs048\pm0\farcs002$ & $132\fdg3\pm0\fdg2$ & $^\mathrm{c}$ \\
2017 Feb 17 & MagAO/Clio2 & $1\farcs056\pm0\farcs005$ & $132\fdg3\pm0\fdg4$ & $^\mathrm{b}$ \\
2019 Apr 07 & VLT/SPHERE & $1\farcs046\pm0\farcs003$ & $133\fdg2\pm0\fdg2$ & $^\mathrm{c}$ \\
\hline
\multicolumn{5}{@{}l@{}}{$^\mathrm{a}$ Data from \citet{Collins2009} with astrometric calibrations from} \\
\multicolumn{5}{@{}l@{}}{\citet{Chauvin2010}; $^\mathrm{b}$ \citet{Wagner2018}; $^\mathrm{c}$ This work.}
\end{tabular}
\end{table}

\begin{figure}
\centering
\resizebox{\hsize}{!}{
\includegraphics{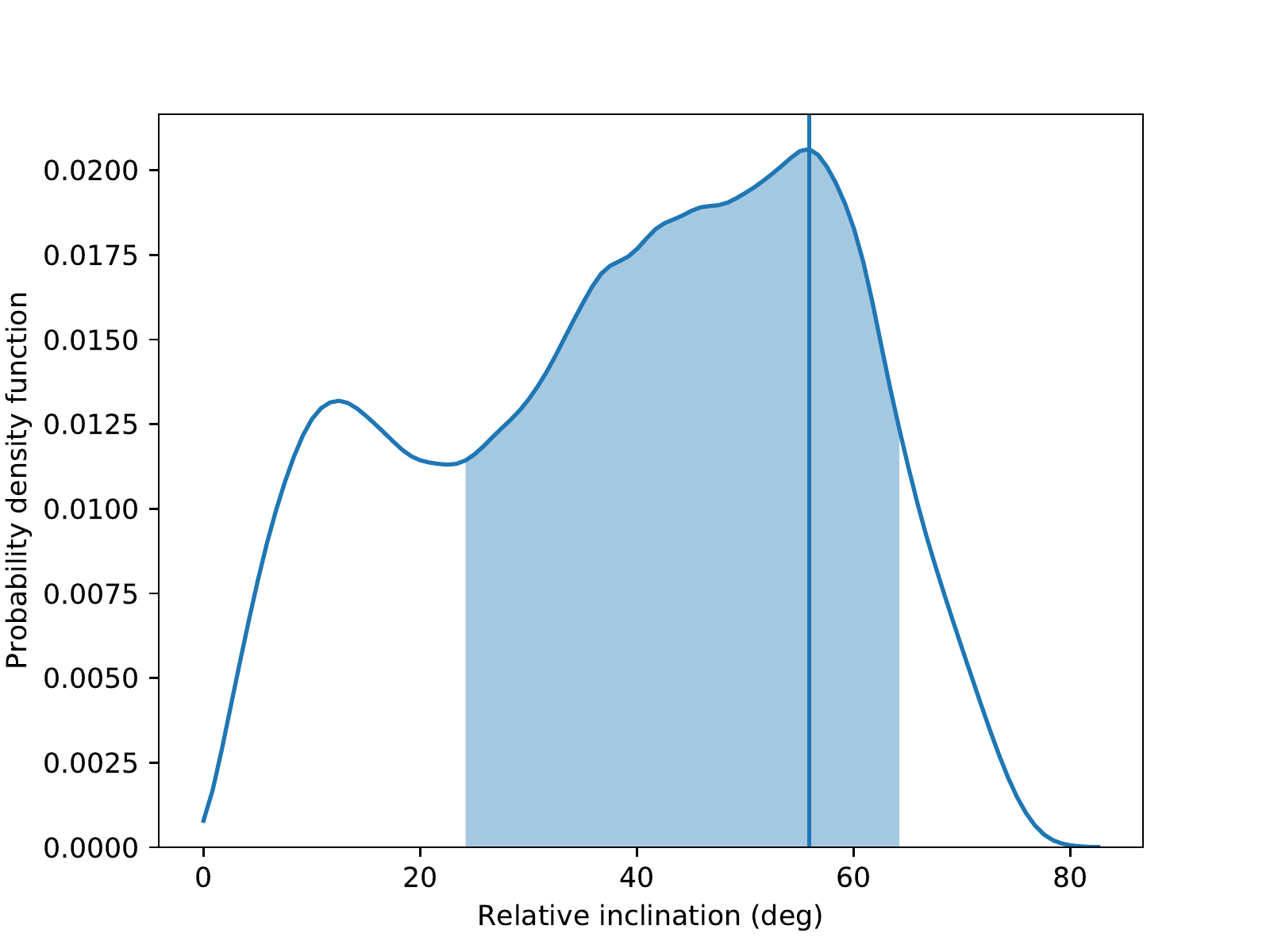}
}
\caption{Probability distribution for $i_\mathrm{rel}$ in the updated MCMC orbital fit. The blue shaded region shows the shortest 68\% confidence interval and the vertical line marks the probability peak.}
\label{Fig:MCMC_irel}
\end{figure}

\begin{figure*}
\centering
\resizebox{\hsize}{!}{
\includegraphics{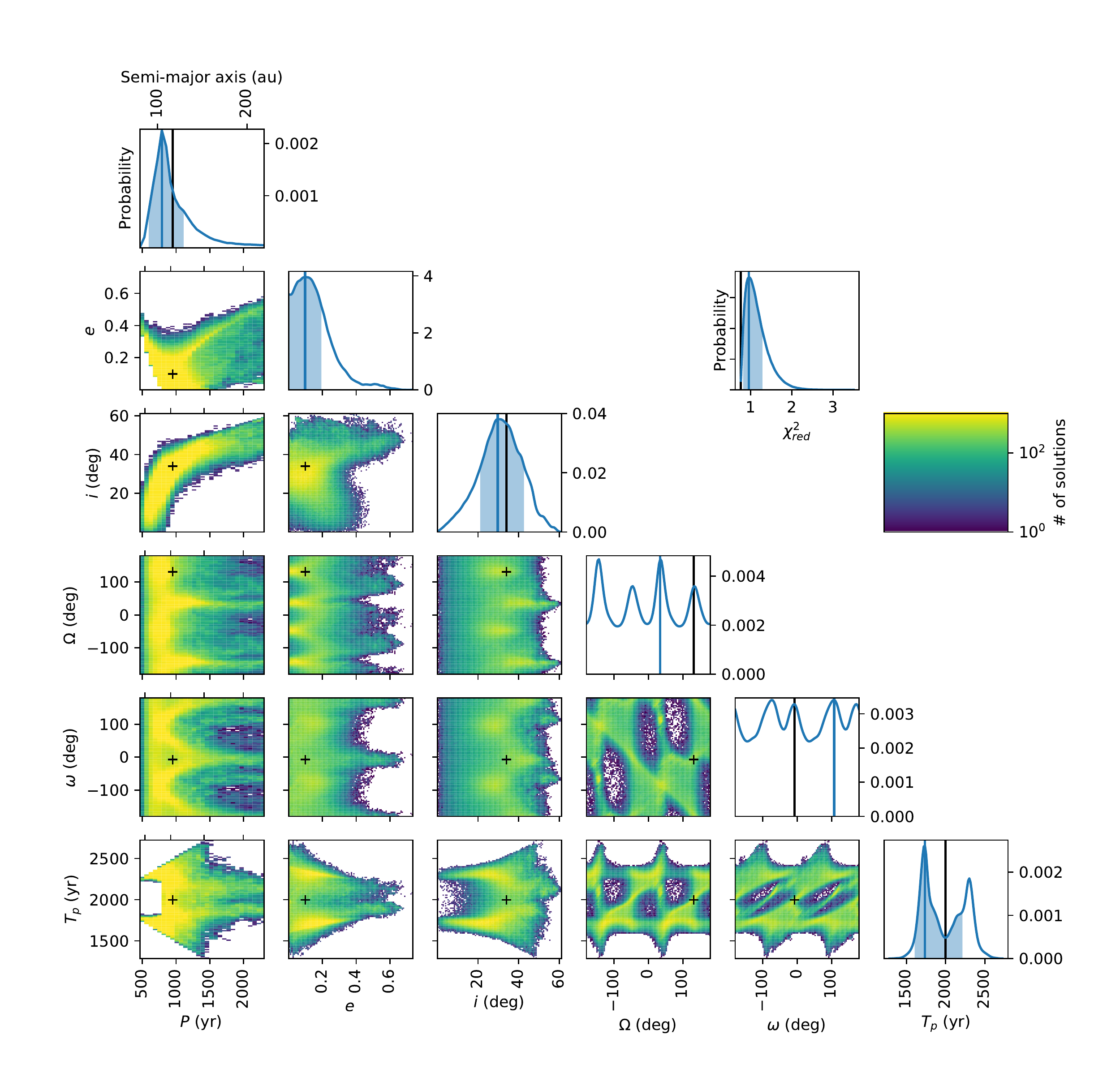}
}
\vspace{-5mm}
\caption{Distribution and correlations of each of the orbital elements fitted by the MCMC algorithm using the updated data. The blue shaded region shows the 68\% confidence interval and the vertical blue line marks the probability peak. The black vertical lines and crosses depict the best fitting orbit (lowest $\chi^2$). The color scale is logarithmic.}
\label{Fig:MCMC_corner}
\end{figure*}

\begin{figure*}
\centering
\resizebox{\hsize}{!}{
\includegraphics{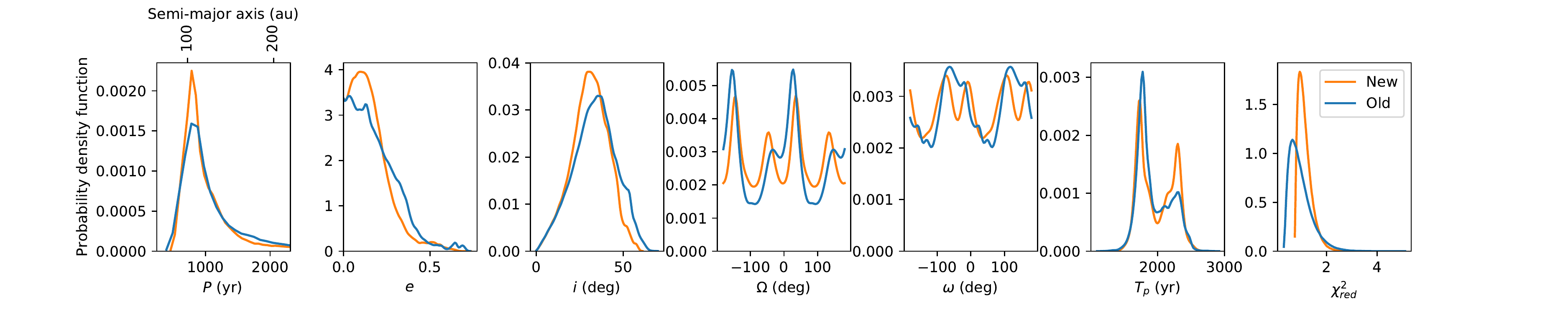}
}
\vspace{-5mm}
\caption{Comparison between the posteriors from the old (blue) and updated (red) astrometry obtained with the MCMC algorithm.}
\label{Fig:MCMC_comp}
\end{figure*}

\begin{figure}
\centering
\resizebox{\hsize}{!}{
\includegraphics{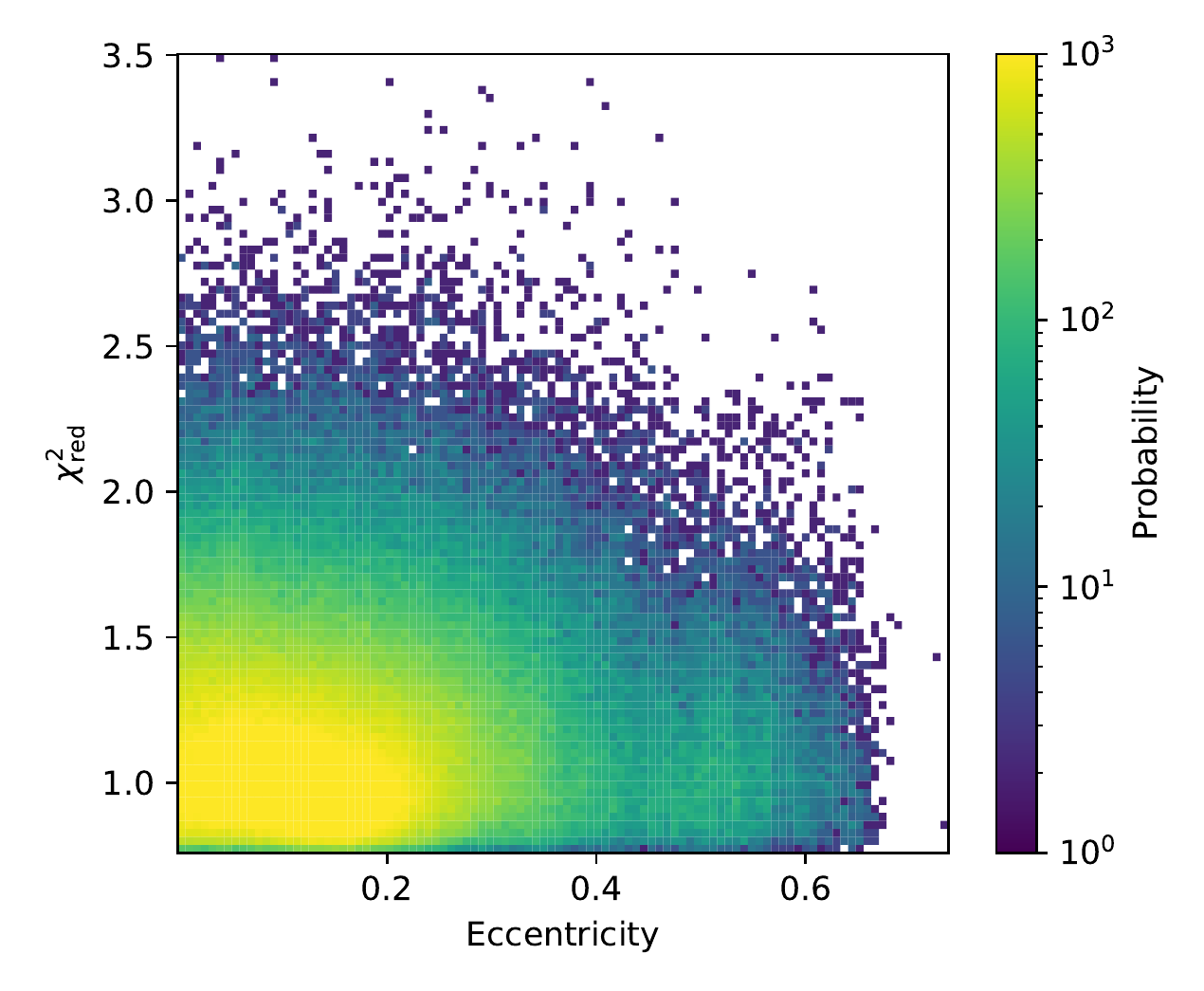}
}
\caption{Probability map in the $\chi_\mathrm{red}^2$~--~eccentricity plane from the updated MCMC orbital fit.}
\label{Fig:MCMC_chi2_e}
\end{figure}

We perform an astrometric fit with this updated catalogue using the same MCMC implementation as \citet{vanderPlas2019} (see also Section~\ref{Sec:MethSetup}). Fig.~\ref{Fig:MCMC_irel} shows the resulting probability distribution for the relative inclination, while those for the 6 orbital elements are displayed in the corner plot of Fig.~\ref{Fig:MCMC_corner}. The confidence intervals remain nearly identical to those obtained in \citet{vanderPlas2019}, as can be seen in the comparison of posteriors plotted in Fig.~\ref{Fig:MCMC_comp}, ensuring the validity of the choice or orbits from Section~\ref{Sec:MethSetup}. In particular, a high relative inclination remains a strong probability. The only significant difference is that the eccentricity probability distribution no longer peaks at zero. A circular orbit is thus no longer the most favoured case. Fig.~\ref{Fig:MCMC_chi2_e} shows that there exists orbits with $e\sim0.3$, like orbit~0, or even much higher, which are very good fits to the data, even though they are not the most probable ones.

\begin{figure*}
\centering
\resizebox{\hsize}{!}{
\includegraphics{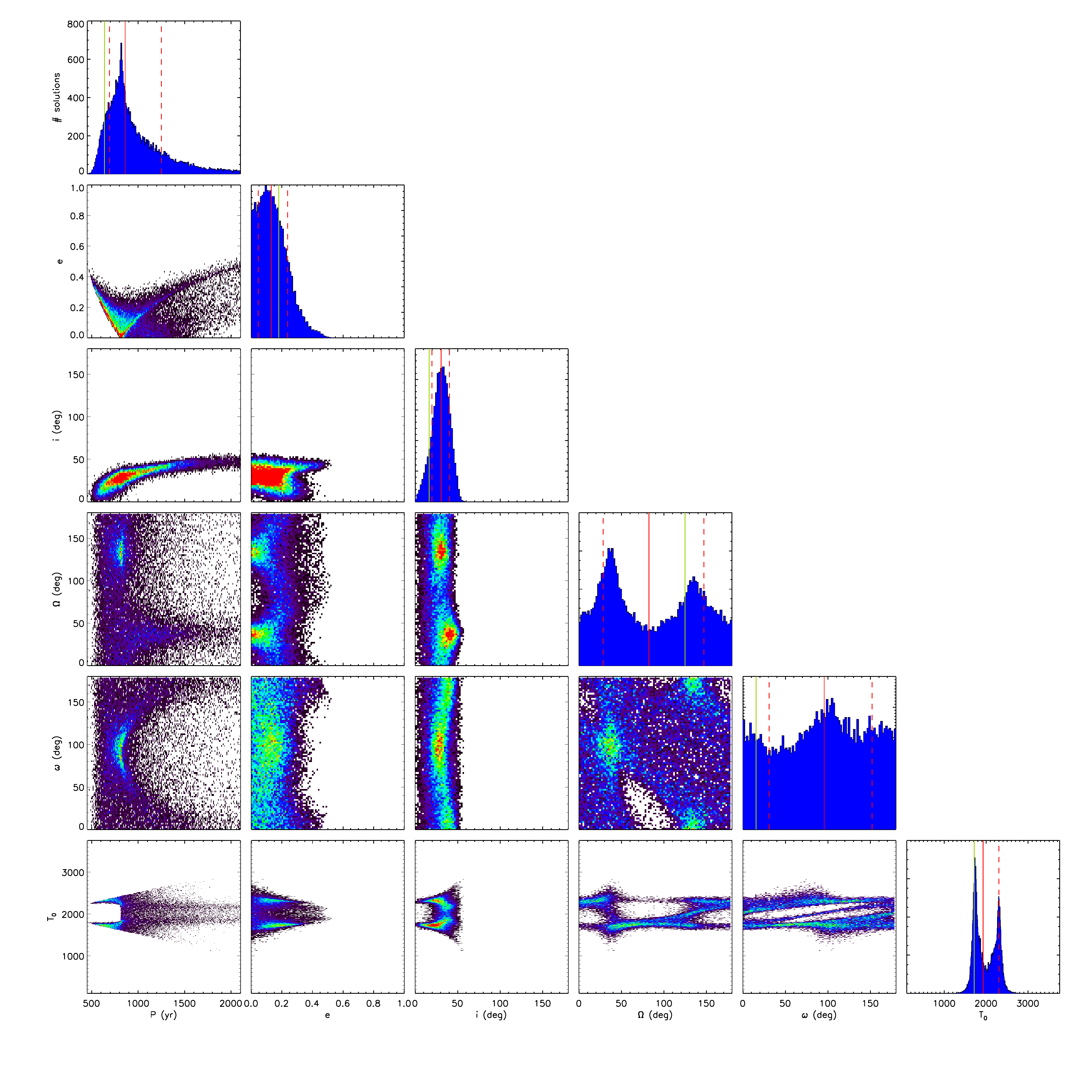}
}
\caption{Same as Fig.~\ref{Fig:MCMC_corner} for the OFTI algorithm.  The dashed red lines delimit the 68\% confidence interval and the solid red lines mark the 50\% percentile. The green lines depict the best fitting orbit (lowest $\chi^2$).}
\label{Fig:OFTI_corner}
\end{figure*}

To validate these new results with an independent method, we use a custom Interactive Data Language (IDL) implementation of the Orbits For The Impatient (OFTI) approach \citep{Blunt2017} described in \citet{Maire2019}. Briefly, we draw random orbits from uniform distributions in $e$, $\cos i$, $\omega$, and $T_0$ (the periastron epoch) and adjust their semi-major axis and longitude of node by scaling and rotating the orbits to match one of the measured astrometric points. The method used to adjust the semi-major axis and longitude of node imposes uniform priors in log\,$P$ (the orbital period) and $\Omega$. Subsequently, the $\chi^2$ probability of each orbit is computed assuming uncorrelated Gaussian errors before performing the rejection sampling test. Fig.~\ref{Fig:OFTI_corner} shows the resulting corner plot of probability distributions for the 6 orbital elements based on 28569 accepted orbits. They are indeed very similar to those found with the MCMC algorithm. In particular, the same eccentricity probability peak at $\sim0.1$ is recovered.

\bsp	
\label{lastpage}
\end{document}